\documentclass[10pt,a4paper]{article}

\usepackage[a4paper,margin=1in]{geometry}
\usepackage{setspace}
\usepackage{amsmath, amssymb}
\usepackage{tikz}
\usetikzlibrary{positioning}
\usepackage{graphicx}
\usepackage{physics}
\usepackage{graphicx}
\usepackage{epstopdf}
\usepackage{hyperref}
\usepackage[nameinlink]{cleveref}
\usepackage{subcaption}
\usepackage{float}
\usepackage{siunitx}
\usepackage{epstopdf}
\usepackage{booktabs}
\graphicspath{
	{Figures/Main/}
	{Figures/Supplementary/}
}
\usepackage{booktabs,tabularx,makecell,array}
\newcolumntype{Y}{>{\raggedright\arraybackslash}X} % ragged-right, auto-wrap
\setcellgapes{2pt}\makegapedcells                   % a little vertical padding

\DeclareGraphicsExtensions{.pdf,.png,.jpg,.jpeg,.eps}

\title{Unified Phase-Field Framework for Antiferroelectric, Ferroelectric and Dielectric Phases: Application to HZO Thin Films}

\author{
	P.~Pankaj$^{1}$,
	Sandeep Sugathan$^{1}$,
	Si~Joon~Kim$^{2}$,
	Pil-Ryung Cha$^{1,*}$
}

\date{}

\begin{document}
	\maketitle
	
	\noindent
	$^{1}$School of Materials Science and Engineering, Kookmin University, Seoul 02707, Republic of Korea\\
	$^{2}$Department of Electrical and Electronics Engineering, Kangwon National University, Chuncheon, Gangwon-Do 24341, Republic of Korea\\[4pt]
	$^{*}$Corresponding author: \texttt{cprdream@kookmin.ac.kr}
	
	\begin{abstract}
	Polycrystalline hafnia-based thin films exhibit mixed ferroelectric (FE), antiferroelectric (AFE), and dielectric (DE) behavior, with switching characteristics strongly influenced by microstructure and phase distribution. Here, we develop a unified grain-resolved three-dimensional phase-field framework for metal–insulator–metal capacitors that simultaneously captures ferroic phase characteristics in realistic polycrystalline microstructures by explicitly incorporating grain topology and crystallographic orientation. Antipolar sublattice kinetics are represented via the coupled evolution of macroscopic and staggered polarization order parameters. All thermodynamic and kinetic parameters are calibrated to experimental $P$--$E$ hysteresis loops and held fixed across all simulations. The results show that phase fractions primarily determine hysteresis character, while vertical segregation of AFE- and FE-rich regions systematically reduces the effective coercive field ($E_c$) under identical electrical loading. Grain-resolved analysis reveals that this reduction arises from microstructure-assisted switching pathways and electrostatic coupling between layers. These findings demonstrate that vertical phase arrangement provides a viable strategy to engineer switching behavior in hafnia-based ferroic capacitors and highlight the importance of explicit microstructural resolution for quantitative phase-field modeling.
    \end{abstract}
    \vspace{6pt}

	\section{Introduction}
	
	Antiferroelectric (AFE) and ferroelectric (FE) materials are actively explored for
	applications spanning nonvolatile memories, negative-capacitance devices, and
	high-power energy-storage capacitors, owing to their rich electric-field-driven phase
	transitions and hysteretic responses
	\cite{Catalan_2015,Liu_2018,Pesic_2016,Randall_Review,Si_Review}.
	In particular, the hallmark double-hysteresis response of AFEs enables large
	recoverable energy density and fast charge--discharge operation, and has also been
	linked to electrocaloric and resistive-memory functionalities
	\cite{Catalan_2015,Liu_2018,Randall_Review}.
	In hafnia-based systems, HfO$_2$- and Hf$_x$Zr$_{1-x}$O$_2$-based thin films have emerged
	as technologically relevant platforms compatible with CMOS processing, exhibiting
	ferroelectric, antiferroelectric-like, and dielectric behavior depending on composition,
	thickness, and processing conditions
	\cite{Boscke_2011,Muller_2012,Park_2015,Si_Review}.
	Notably, AFE-like ZrO$_2$/HfO$_2$ stacks have been proposed to mitigate the voltage-scaling
	and cycling-lifetime limitations associated with high-$E_c$ FE hafnia, while enabling
	memory-relevant switching concepts and high-efficiency energy storage
	\cite{Pesic_2016}. Together, these studies highlight the central challenge in antiferroelectric and
	ferroelectric device design: achieving reduced switching fields and controlled
	hysteresis characteristics without sacrificing reversibility or endurance.

	In hafnia-based films, AFE-like switching is often interpreted primarily in terms of
	electric-field-induced crystallographic phase transitions between nonpolar and
	polar phases, with the critical switching field governed by relative phase
	energetics rather than explicit antipolar domain ordering
	\cite{Weng_2023,Yan_2025}. In contrast, studies on Pb-based perovskite
	antiferroelectrics have demonstrated that deliberately disrupting long-range
	antipolar order—through entropy-driven disorder, grain refinement, antipolar
	frustration, or engineered polarization heterogeneity—can diffuse the AFE–FE
	transition, reshape hysteresis behavior, and enhance functional performance
	\cite{Gao_2016,Liu_Xu_2020,Zhou_2025,Yang_2025}. In such systems, depolarization
	fields and electrostatic boundary conditions play a central role in determining
	the reversibility and stability of field-induced phase transitions
	\cite{Geng_2018}.
	
	Experimentally, hafnia-based thin films exhibit a wide spectrum of switching
	responses, ranging from FE-like single hysteresis loops to AFE-like double loops
	and nearly linear dielectric behavior, reflecting strong sensitivity to phase
	fraction, microstructure, and cycling history \cite{Park_2015,Muller_2012}.
	Notably, polarization cycling in Hf$_x$Zr$_{1-x}$O$_2$ films can induce a systematic
	evolution of AFE hysteresis, in which initially separated sub-loops progressively
	merge into a single loop with finite remanent polarization while retaining
	exceptionally high endurance \cite{Chen_2022}. From a device perspective, these
	observations reveal a fundamental trade-off in hafnia-based memories: FE-rich
	films provide finite $P_r$ but typically require relatively high switching fields
	and exhibit pronounced cycling sensitivity, whereas AFE-rich films exhibit highly
	reversible switching with near-zero $P_r$. This trade-off motivates strategies to
	reduce the effective coercive field $E_c$ while preserving FE-like hysteresis
	characteristics, without relying on chemistry- or defect-driven phase
	stabilization.
	
	Phase-field modeling has played a central role in elucidating FE and AFE switching
	mechanisms, including domain formation, hysteresis evolution, and phase
	competition. Unified multi-domain phase-field frameworks capable of simulating
	coexisting FE, AFE, and dielectric regions in hafnia-based films have been reported
	and successfully used to reproduce mixed hysteresis behavior and cycling trends
	\cite{Chang_2022,Islam_2025}. However, in most such models, polycrystallinity is
	represented through grid-wise phase mixtures or reduced-dimensional geometries,
	rather than an explicit three-dimensional (3-D) grain-resolved microstructure. As a
	result, these approaches are limited in their ability to isolate how realistic
	grain connectivity, vertical phase arrangement, and grain-scale electrostatic
	coupling govern the practical trade-off between reducing $E_c$ and retaining
	FE-like $P_r$ in polycrystalline capacitors. At the same time, experimental and
	atomistic studies suggest that confinement-driven grain energetics and
	interface-related effects strongly influence phase stability in ZrO$_2$/HZO,
	promoting stabilization of the orthorhombic phase and reshaping the free-energy
	landscape under applied electric fields \cite{Boscke_2011,Muller_2012,Park_2017,Ganser_2024}.
	
	In this work, we address these gaps by developing a unified grain-resolved 3-D phase-field model for polycrystalline hafnia-based thin films. The model explicitly resolves polycrystallinity using a 3-D Voronoi grain topology with layer-resolved vertical phase assignment and evolves both $P_m$ and $P_s$, enabling direct representation of antipolar sublattice kinetics. A uniaxial switching formulation appropriate for metal–insulator–metal capacitors is employed, in which nonlinear polarization instabilities are confined to the film-normal direction while depolarization fields are treated self-consistently through Poisson coupling with phase-dependent dielectric permittivity. This framework enables systematic investigation of how vertical phase arrangement and grain-scale electrostatic coupling influence switching pathways and control $E_c$ and $P_r$ in polycrystalline hafnia films under identical electrical loading.
	
\section{Phase-field modeling framework and thermodynamic description}
\label{sec:formulation}
	
	We employ a 3-D phase-field framework to simulate electric-field-driven
	switching in hafnia-based polycrystalline thin films containing AFE, FE, and DE grains.
	Polycrystallinity is represented by a fixed 3-D grain structure in which each grain is assigned a phase identity and a crystallographic orientation specified by the ZXZ Euler-angle convention; details of microstructure generation are provided in Section~\ref{sec:PolyX}. The total free-energy of the polycrystalline film is given by
	\begin{equation}
		F=\int_V\left(
		f_{\mathrm{bulk}}
		+f_{\mathrm{grad}}
		+f_{\mathrm{elc}}
		+f_{\mathrm{elas}}
		+f_{\mathrm{ap}}
		\right)\,dV .
	\end{equation}
	where $f_{\mathrm{bulk}}$ describes the thermodynamics of AFE, FE, and DE grains,
	$f_{\mathrm{grad}}$ penalizes spatial variations of the polarization fields,
	$f_{\mathrm{elc}}$ accounts for long-range electrostatic interactions and depolarization
	fields, $f_{\mathrm{elas}}$ represents elastic energy, and $f_{\mathrm{ap}}$ represents coupling to the applied electric field.

	\subsection{Bulk thermodynamics and gradient energy}
	
	The bulk thermodynamics is described using the Kittel two-sublattice model~\cite{Kittel_1951,Rabe_Wiley_2012},
	\begin{equation}
		f_{\mathrm{bulk}}(P_a,P_b)=
		\frac{\alpha(T)}{2}(P_a^2+P_b^2)
		+\frac{\beta}{4}(P_a^4+P_b^4)
		+\frac{\gamma}{6}(P_a^6+P_b^6)
		+g\,P_a P_b ,
	\end{equation}
	where $\alpha, \beta, \gamma, g$ are the Landau coefficients, $P_a$ and $P_b$ denote the two-sublattice polarization. Introducing the macroscopic and staggered order parameters $P_m=P_a+P_b$ and $P_s=P_a-P_b$, and assigning material parameters on a grain-by-grain basis, with $\mathcal{P}(k)\in\{\mathrm{AFE},\mathrm{FE},\mathrm{DE}\}$ denoting the phase identity of grain $k$, the bulk free-energy density is expressed in the local crystallographic frame of grain $k$ as~\cite{Lum_2022}
	
	\begin{equation}
		\begin{aligned}
			f_{\mathrm{bulk}}^{L,k}(\mathbf{P}_m,\mathbf{P}_s) &=
			\frac{\alpha^{\mathcal{P}(k)}(T)}{4}\!\left[(P_m^L)^2+(P_s^L)^2\right]
			+\frac{\beta^{\mathcal{P}(k)}}{32}\!\left[(P_m^L)^4
			+6(P_m^L)^2(P_s^L)^2+(P_s^L)^4\right] \\
			&\quad
			+\frac{\gamma^{\mathcal{P}(k)}}{192}\!\left[(P_m^L)^6
			+15(P_m^L)^4(P_s^L)^2
			+15(P_m^L)^2(P_s^L)^4
			+(P_s^L)^6\right]
			+\frac{g^{\mathcal{P}(k)}}{4}\!\left[(P_m^L)^2-(P_s^L)^2\right],
		\end{aligned}
		\label{eq:FE_Pm_Ps_L_grain}
	\end{equation}
	where the superscript $L$ denotes the local crystallographic frame and
	$\alpha(T)=g+\alpha_0(T-T_0)$. Within this formulation, AFE, FE, and paraelectric states
	correspond to $P_m=0$, $P_s=0$, and $P_m=P_s=0$, respectively~\cite{Kittel_1951,Cross_1967}.
	
Experimental studies on orthorhombic HZO and ZrO$_2$ thin films under metal–insulator–metal capacitor boundary conditions show that polarization switching is predominantly confined to the film-normal direction~\cite{Boscke_2011,Muller_2012}. Although the film is 3-D and polycrystalline, texture analysis and wake-up-induced polar-axis reorientation indicate a primarily out-of-plane ferroelectric response~\cite{Lederer_2019}.
	Consequently, only the z-component of polarization is treated as switchable in this work, whereas transverse components are treated as non-switching contributions through an effective local dielectric stiffness $\chi_f$. Suppressing the explicit grain index for notational
	simplicity, the local bulk free-energy density reduces to
	\begin{equation}
		\begin{aligned}
			f_{\mathrm{bulk}}^{L}(\mathbf{P}_m,\mathbf{P}_s) &=
			\frac{\alpha(T)}{4}\!\left[(P_{m,z}^{L})^2+(P_{s,z}^{L})^2\right]
			+\frac{\beta}{32}\!\left[(P_{m,z}^{L})^4
			+6(P_{m,z}^{L})^2(P_{s,z}^{L})^2
			+(P_{s,z}^{L})^4\right] \\
			&\quad
			+\frac{\gamma}{192}\!\left[(P_{m,z}^{L})^6
			+15(P_{m,z}^{L})^4(P_{s,z}^{L})^2
			+15(P_{m,z}^{L})^2(P_{s,z}^{L})^4
			+(P_{s,z}^{L})^6\right]
			+\frac{g}{4}\!\left[(P_{m,z}^{L})^2-(P_{s,z}^{L})^2\right] \\
			&\quad
			+\frac{1}{2\chi_f}\sum_{i=x,y}
			\left[(P_{m,i}^{L})^2+(P_{s,i}^{L})^2\right].
		\end{aligned}
		\label{eq:FE_Pm_Ps_xyz_L}
	\end{equation}
	
	The gradient energy density is written in the global frame as
	\begin{equation}
		f_{\mathrm{grad}}^{G}(\mathbf{P}_m)
		=
		\frac{G_{11}}{2}
		\sum_{i=x,y,z}
		\left[
		(\nabla P_{m,i}^{G})^{2}
		+
		(\nabla P_{s,i}^{G})^{2}
		\right],
	\end{equation}
	where $G_{11}$ denotes isotropic gradient-energy coefficient, the superscript $G$ denotes the global crystallographic frame. 
	
	\subsection{Electrostatics and applied-field coupling}
	
	Electrostatic interactions arising from spatial variations of the macroscopic polarization are described by the energy density
	\begin{equation}
		f_{\mathrm{elc}}^{G}(\mathbf{P}_m,\phi) = -\tfrac{1}{2}\,E_i^{G} P_{m,i}^{G},
	\end{equation}
	where the internal electric field is defined by
	$\mathbf{E}^G=-\nabla\phi^G$, with $\phi^G$ denoting the electrostatic
	potential. In the absence
	of free charges, electrostatic equilibrium requires $\nabla\!\cdot\!\mathbf{D}^G=0$, which
	leads to the Poisson equation
	\begin{equation}
		(\varepsilon_{ij}\phi_{,j}^G)_{,i}=\frac{P_{m,i,i}^G}{\varepsilon_0}.
		\label{eq:PoissonEq}
	\end{equation}
	Here, $\varepsilon_{ij}(\mathbf{r})=\varepsilon_r^{\alpha}\delta_{ij}$ denotes the
	phase-dependent isotropic relative permittivity, and $\varepsilon_0$ is the vacuum
	permittivity. Eq.~\eqref{eq:PoissonEq} is solved using a conjugate-gradient method with
	a Jacobi preconditioner, subject to periodic in-plane and Dirichlet boundary conditions along the film normal.
	
	The parameter $\chi_f$ represents a local dielectric stiffness entering only through the
	bulk free-energy functional as a harmonic penalty on polarization, whereas the relative
	dielectric permittivity tensor $\varepsilon_{ij}$ appears only in Poisson’s equation
	and governs long-range electrostatic screening. These contributions are therefore treated independently to avoid double counting of dielectric response, as discussed in analyses separating bulk and depolarization-field contributions to small-signal capacitance~\cite{Koduru_2025}.

	The energetic contribution of an externally applied electric field is
	\begin{equation}
		f_{\mathrm{ap}}^{G}(\mathbf{P}_m) = - E_{\mathrm{ap,i}}^{G} P_{m,i}^{G},
		\label{eq:app_field}
	\end{equation}
	which reduces to $f_{\mathrm{ap}}^{G}=-E_{\mathrm{ap,z}}^{G}P_{m,z}^{G}$ under out-of-plane
	capacitor loading. Field-induced modification of AFE order occurs indirectly through
	$P_{m,z}$ rather than via a direct linear coupling to the staggered order
	parameter~\cite{Vakhrushev_2021}.

Elastic energy is neglected in the present formulation as our focus is to isolate electrostatic, microstructural, and phase-fraction effects on polarization switching in polycrystalline hafnia thin films. The Landau coefficients are calibrated directly against experimental $P$–$E$ hysteresis loops of comparable thin-film geometry (see Section~\ref{subsec:parameter_calibration}); thus, the resulting effective bulk free-energy landscape implicitly reflects the average elastic constraints associated with substrate clamping and processing-induced residual stresses. Also, previous phase-field studies indicate that elastic coupling exerts only a secondary influence on the macroscopic hysteresis response relative to electrostatic interactions and domain-wall energetics~\cite{Sugathan,KEVIN}.

	\subsection{Phase-transition thermodynamics}
	
The AFE--FE transition in HZO-based systems is dominated by polarization
along the polar crystallographic axis. Transverse components respond
predominantly linearly to the applied field and are incorporated through
an effective dielectric stiffness $\chi_f$. Accordingly, only the polar
($z$-axis) components are retained in describing the transition
thermodynamics. Under this approximation, Eq.\eqref{eq:FE_Pm_Ps_xyz_L} can be expressed in terms of an
	effective scalar order parameter $P$:
	\begin{equation}
		\mathrm{F}(P) = \frac{1}{2} a(T) P^2 + \frac{1}{4} b P^4 + \frac{1}{6} c P^6,
		\label{eq:Landau_gen}
	\end{equation} 
	where $P$ denotes $P_{s,z}^L$ for the AFE mode or $P_{m,z}^L$ for the FE mode,
	with $b=\beta/8$ and $c=\gamma/32$.
	The temperature-dependent quadratic coefficients are
	\begin{align}
		a_s(T) &= \frac{\alpha(T)-g}{2}
		= \frac{\alpha_0}{2}(T-T_0), \quad \text{AFE}, \nonumber\\
		a_m(T) &= \frac{\alpha(T)+g}{2}
		= \frac{\alpha_0}{2}\!\left(T-T_0+\frac{2g}{\alpha_0}\right), \quad \text{FE},
		\label{eq:a_T_relation}
	\end{align}
	yielding instability temperatures $T_s^*=T_0$ and $T_m^*=T_0-2g/\alpha_0$.
	The coupling parameter $g$ therefore shifts the FE instability downward relative to
	the AFE branch.
	
\begin{figure}[htbp]
		\centering
		\includegraphics[width=8.5cm]{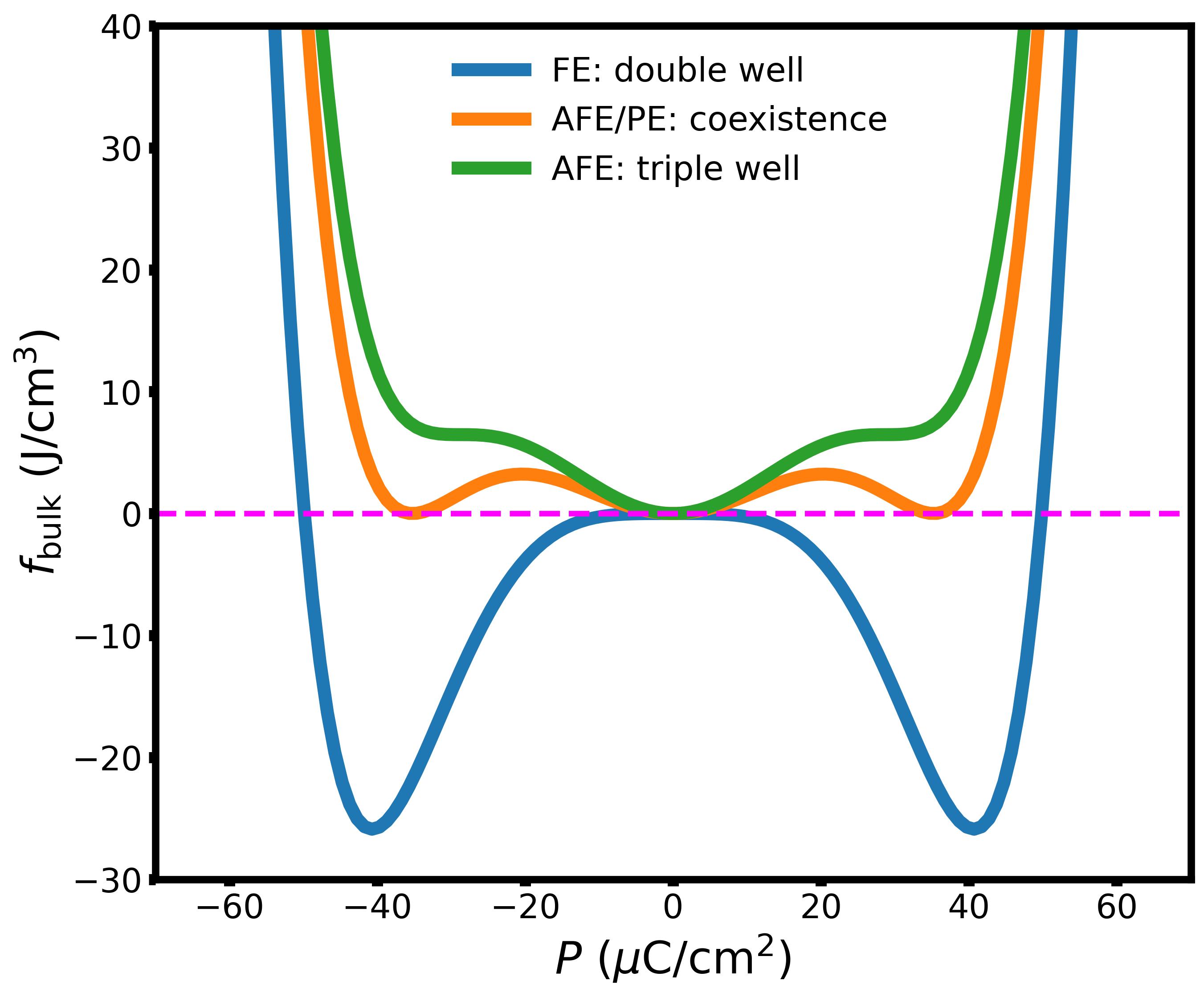}
		\caption{
			Zero-field bulk free-energy density $f(P)$ versus polarization $P$ for the
			scalar Kittel--Landau model.
			Curves correspond to FE double-well ($a<0$),
			AFE/PE coexistence ($a=a_{\mathrm{coex}}$), and
			AFE triple-well ($a_{\mathrm{ord}}^{spn}>a>a_{\mathrm{coex}}$).
		}
		\label{fig:F_P_gen}
	\end{figure}

	At high temperature ($a>0$), the system is paraelectric (PE) with a single minimum at $P=0$.
	Upon cooling, the free-energy landscape evolves into a
	\emph{triple-well} form, marked by the appearance of two metastable polar minima while the
	central paraelectric minimum remains globally stable. This is characterized by the \emph{ordered spinodal}, $a_{ord}^{spn}=b^2/(4c)$. At coexistence,
	$a_{coex}=3b^2/(16c)$, where paraelectric and polar states are degenerate. Further cooling removes the central minimum at the disordered spinodal $a=0$, leaving
	a stable double-well profile. The free-energy landscapes are shown in Fig.~\ref{fig:F_P_gen}.
	
	%Here, the term spinodal is used in its thermodynamic sense to denote the loss of
	%local stability of a free-energy extremum, independent of the specific kinetic
	%pathway by which the system subsequently evolves.
	
The presence of an intermediate triple-well free-energy landscape provides the fundamental thermodynamic distinction between AFE double hysteresis and FE single hysteresis: in AFE systems, a low-field nonpolar minimum enables reversible field-induced switching, whereas FE systems exhibit a stable double-well potential supporting spontaneous polarization. This interpretation is consistent with Landau stability analyses and Gibbs free-energy
	reconstructions of AFE switching~\cite{Lum_2022,Lomenzo2020,Segatto2023}. Pulsed measurements in ZrO$_2$ further suggest that AFE–FE switching dynamically traverses spinodal-bounded unstable regions~\cite{Hoffmann2022}, corresponding to loss of local stability of the uniform polarization state~\cite{Ke_2020}.

	For $g>0$, the AFE branch softens first, producing a triple-well
	landscape and characteristic AFE double hysteresis.
	For $g<0$, the FE branch dominates, and the system exhibits a spontaneous double-well
	state and single hysteresis loop.
	Accordingly, the dominant instability pathway is
	\begin{equation}
		a(T) =
		\begin{cases}
			a_s(T), & g>0 \quad (\text{AFE-dominated}),\\
			a_m(T), & g<0 \quad (\text{FE-dominated}).
		\end{cases}
	\end{equation}
	
Experimental studies have reported that electric-field-induced AFE–FE switching is first-order, characterized by discontinuous polarization and hysteresis~\cite{Tan_2010}.
	To obtain triple-well regime in second-order formulation, it is required to eliminate the antipolar order parameter via its stationary
	condition~\cite{Chang_2022}. However, this enforces instantaneous equilibration and removes it as an independent field. Consequently, finite-rate relaxation, spatial pinning, and domain-wall energetics of $P_s$ cannot be captured, and the kinetic pathways by which antipolar order evolves during field-driven switching remain inaccessible. The need for explicit evolution of $P_s$ is reinforced by sublattice-resolved analyses of polarization switching in hafnia-based AFE~\cite{Kim_2023}. We therefore adopt
	a first-order free-energy description with explicit antipolar kinetics, consistent
	with classical Pb-based AFE systems~\cite{Tan_2010,Li_2023}.

	\subsection{Kinetic evolution}
	
	While $f_{\mathrm{bulk}}^{L}$ is evaluated in the local crystallographic frame of each grain, all other energy contributions are evaluated in the global frame. The two frames are related by $\mathbf{P}^L=\mathbf{R}\mathbf{P}^G$:
	\begin{equation}
		\frac{\delta F}{\delta \mathbf{P}^G}
		=
		\mathbf{R}^{\mathsf T}
		\frac{\delta F}{\delta \mathbf{P}^L}.
	\end{equation}
	Here, $\mathbf{R}$ denotes the orthogonal rotation matrix that transforms polarization vectors from the global to the local crystallographic frame of each grain.
	
	The domain evolution within each grain is described by TDGL equations:
	\begin{equation}
		\frac{\partial P_{m,i}^G}{\partial t} = -\Gamma \,\frac{\delta F}{\delta P_{m,i}^G},
		\qquad
		\frac{\partial P_{s,i}^G}{\partial t} = -\Gamma \,\frac{\delta F}{\delta P_{s,i}^G},
		\quad i\in\{x,y,z\}.
	\end{equation}
	where $\Gamma$ is the kinetic coefficient.
	
	All governing equations are expressed in nondimensional form using characteristic reference scales; details are provided in Section 1 of the Supplementary Information.
	
	\section{Experimental methods and parameter calibration}

	\subsection{Experimental method}
	Ferroelectric TiN/Hf$_{0.5}$Zr$_{0.5}$O$_2$/TiN and antiferroelectric
	TiN/ZrO$_2$/TiN capacitors were fabricated on p-type Si substrates.
	A 300-nm-thick SiO$_2$ layer was thermally grown on the substrates prior
	to device fabrication. TiN bottom electrodes ($\sim$90 nm) were deposited
	at room temperature by RF sputtering. Hf$_{1-x}$Zr$_x$O$_2$ ($x = 0.5, 1$)
	thin films were deposited by atomic layer deposition (ALD) at 250~$^\circ$C
	using Hf[N(CH$_3$)$_2$]$_4$ (TDMA-Hf), Zr[N(CH$_3$)$_2$]$_4$ (TDMA-Zr),
	and O$_3$ as the Hf precursor, Zr precursor, and oxygen source, respectively, with a growth
	rate of $\sim$0.2 nm/cycle. The Zr composition was controlled by the ALD cycle ratio, and the total number of cycles was adjusted to achieve a uniform film thickness of
	$\sim$10 nm. After Hf$_{1-x}$Zr$_x$O$_2$ deposition, TiN top electrodes
	($\sim$90 nm) were deposited by RF sputtering under the same conditions
	as the bottom electrodes. The samples were then annealed by rapid thermal
	annealing (RTA) at 400~$^\circ$C for 60 s under N$_2$ atmosphere. TiN top
	electrodes were patterned using photolithography and wet etching with a
	Pd/Au hard mask. Electrical properties were characterized by $P$--$E$ hysteresis measurements using a semiconductor parameter analyzer
	(Keithley 4200A-SCS) equipped with a pulse measurement unit (4225-PMU) at
	10 kHz. Prior to measurement, all devices were subjected to $10^5$ wake-up
	cycles under an electric field of 3 MV/cm to stabilize the FE and AFE
	responses.

	\subsection{Parameter Calibration}
	\label{subsec:parameter_calibration}

For the simulations, experimentally relevant electric-field loading conditions are reproduced using a triangular electric-field waveform with frequency $\nu = 10~\mathrm{kHz}$ and peak amplitude $3~\mathrm{MV/cm}$ applied across a $10~\mathrm{nm}$-thick film. Additional numerical details of the sweep-rate implementation, including discretization parameters and timestep considerations, are provided in Section 2 of the Supplementary Information.
    
\begin{figure}[ht]
		\centering
		\begin{subfigure}[b]{0.48\linewidth}
			\centering
			\includegraphics[width=\linewidth]{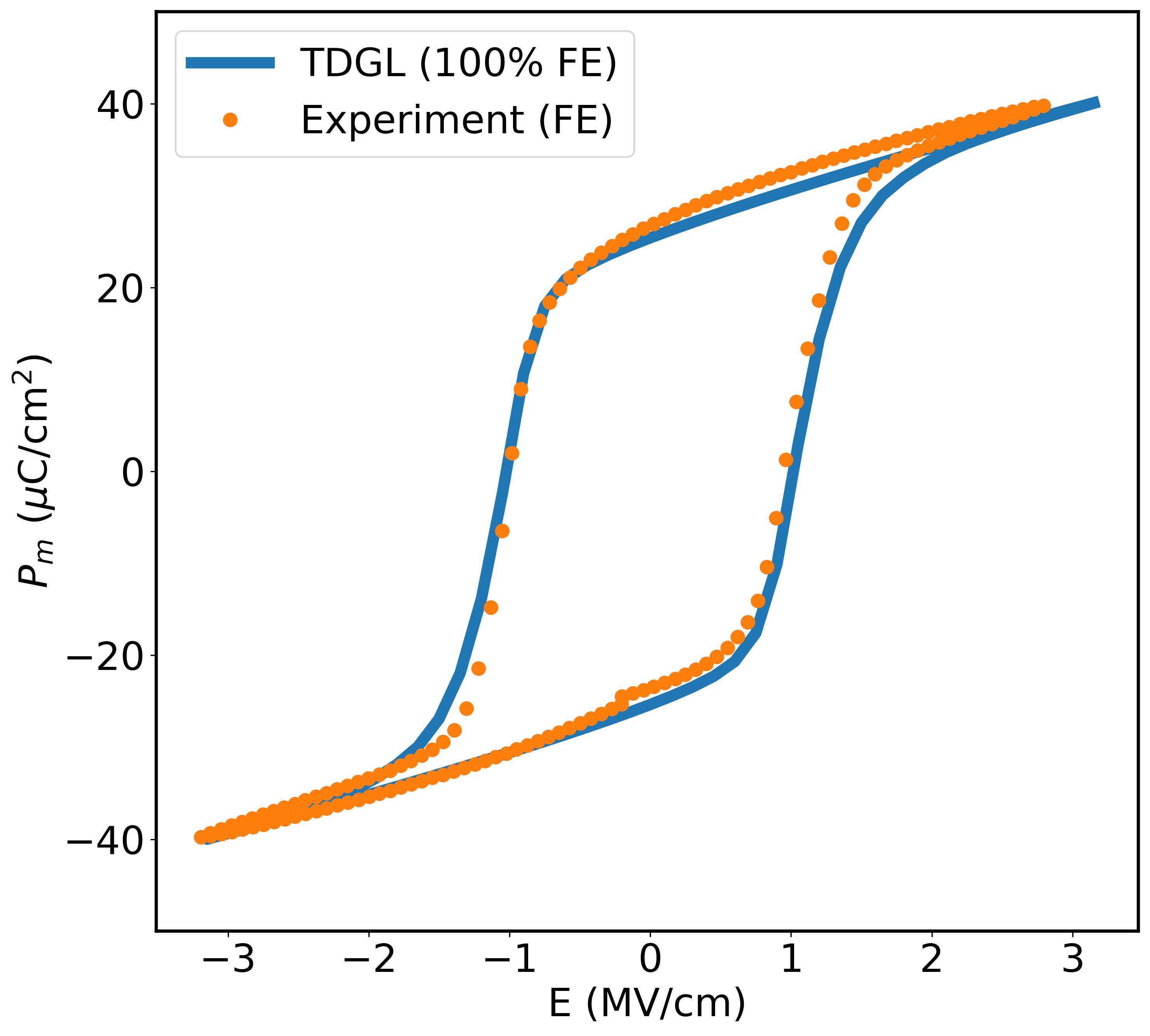}
			\caption{Ferroelectric (HZO): calibrated TDGL vs experiment.}
			\label{fig:calib_FE}
		\end{subfigure}
		\hfill
		\begin{subfigure}[b]{0.48\linewidth}
			\centering
			\includegraphics[width=\linewidth]{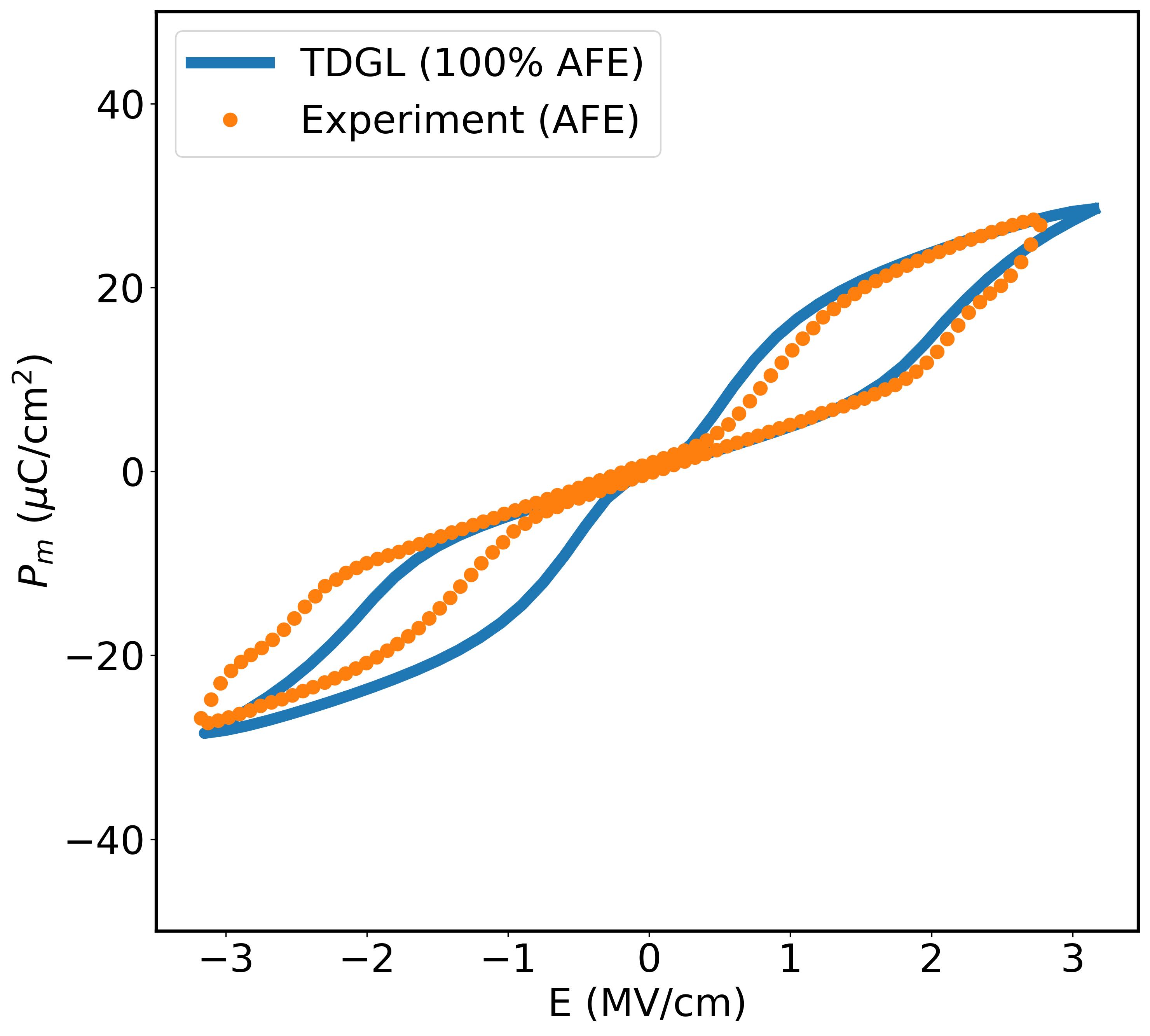}
			\caption{Antiferroelectric (ZrO$_2$): calibrated TDGL vs experiment.}
			\label{fig:calib_AFE}
		\end{subfigure}
		\caption{Calibration of thermodynamic and kinetic parameters using experimental
			$P$--$E$ hysteresis loops of polycrystalline thin films. Solid lines denote TDGL
			simulations and dashed lines experimental data, obtained under identical
			electric-field loading conditions.}
		\label{fig:calibration_loops}
	\end{figure}

	The Landau coefficients and selected kinetic parameters are calibrated against
	experimental $P$--$E$ hysteresis loops measured on polycrystalline HZO and
	ZrO$_2$ thin films, while the remaining kinetic parameters are
	adopted from established literature. Fig.~\ref{fig:calibration_loops} compares the calibrated TDGL responses with the corresponding experimental data, and the fitted
	coefficients are summarized in Table~\ref{tab:landau_params}. Detailed analysis
	of microstructural domain evolution is deferred to the polycrystalline
	simulations discussed in Section~\ref{sec:1L_microstrucutres}. The calibrated parameters, electric-field increments, and temporal discretization are treated as intrinsic and kept fixed for all simulations. Next, we first establish the ideal AFE single-crystal thermodynamic limit from equilibrium solutions and then introduce spatially uniform TDGL kinetics to elucidate the origin of double hysteresis. Finally, the framework is extended to fully 3-D grain-resolved polycrystalline films to examine the influence of single-layer and multilayer architectures on $E_c$ and $P_r$.

	\begin{table}[ht]
		\centering
		\caption{Calibrated material, kinetic, and geometric parameters.}
		\label{tab:landau_params}
		\vspace{2mm}
		
		\setlength{\tabcolsep}{6pt}
		
		\makebox[\linewidth][c]{%
			\begin{tabular}{l S S S}
				\hline
				Parameter & {AFE (ZrO$_2$)} & {FE (HZO)} & {DE$^{\dagger}$} \\
				\hline
				
				$\alpha$ ($\mathrm{J\,m\,C^{-2}}$)
				& \num{5.022e8}
				& \num{-2.98e8}
				& \num{9.5e9} \\
				
				$\beta$ ($\mathrm{J\,m^{5}\,C^{-4}}$)
				& \num{-9.065e10}
				& \num{-8.520e9}
				& 0 \\
				
				$\gamma$ ($\mathrm{J\,m^{9}\,C^{-6}}$)
				& \num{2.190e12}
				& \num{3.134e11}
				& 0 \\
				
				$g$ ($\mathrm{J\,m\,C^{-2}}$)
				& \num{5.453e8}
				& \num{-2.484e8}
				& 0 \\
				
				$\chi_f$ ($\mathrm{C^{2}\,J^{-1}\,m^{-1}}$)
				& \num{5.0e-10}
				& \num{5.0e-10}
				& \num{2.1e-10}\textsuperscript{\scriptsize\cite{Sugathan}} \\

				$G_{11}$ ($\mathrm{J\,m^{3}\,C^{-2}}$)
				& \num{1.0e-9}
				& \num{5.066e-10}\textsuperscript{\scriptsize\cite{Sugathan}} 
				& \num{2.5e-9} \\
				
				$\Gamma$ ($\mathrm{C^{2}\,J^{-1}\,m^{-1}\,s^{-1}}$)
				& \multicolumn{1}{c}{0.2}
				& \multicolumn{1}{c}{$0.833$\textsuperscript{\scriptsize\cite{Koduru_JAP_2023}}}
				& \multicolumn{1}{c}{1} \\

				$\varepsilon_r$
				& \multicolumn{1}{c}{40\textsuperscript{\scriptsize\cite{Lomenzo_AEM_2022,Park_APL_2014}}}
				&\multicolumn{1}{c}{30\textsuperscript{\scriptsize\cite{KEVIN, Lomenzo_AEM_2022,Park_APL_2014}}}
				&\multicolumn{1}{c}{15\textsuperscript{\scriptsize\cite{Lomenzo_AEM_2022,Park_APL_2014}}}
				\\
				
				\hline
				
				\multicolumn{4}{c}{\textit{Geometric and microstructural parameters}} \\
				\hline
				
				Thin-film surface area ($n_x \times n_y$)
				& \multicolumn{3}{c}{940 $\times$ 940 nm$^{2}$} \\
				
				Thin-film thickness ($t_z$)
				& \multicolumn{3}{c}{10 nm} \\
				
				Number of grains
				& \multicolumn{3}{c}{800} \\
				
				$\Delta h$
				& \multicolumn{3}{c}{1 nm} \\
				
				$\Delta t$
				& \multicolumn{3}{c}{0.06 ns} \\
				\hline
			\end{tabular}
		}
		
		\vspace{2mm}
		\raggedright
		\footnotesize
		\footnotesize
		$^{\dagger}$ For the DE phase, the parameters $\alpha$ and $\chi_f$ are chosen such that
		$\alpha/4 = 1/(2\chi_f)$, yielding an isotropic linear dielectric response with no
		spontaneous polarization. Higher-order Landau coefficients
		($\beta$, $\gamma$, $g$) are therefore set to zero.
		
	\end{table}

	\section{Results and discussion}
	
	\subsection{Intrinsic switching response in single-crystal film}
	\label{sec:SingleX}
In this section, we analyze the ideal single-crystal response under quasi-static electric-field loading to establish the intrinsic thermodynamic switching pathway prior to introducing kinetic effects. Equilibrium polarization states are obtained by minimizing Eq.~\eqref{eq:FE_Pm_Ps_L_grain} using a Newton–Raphson scheme:
	\begin{equation}
		\frac{\partial f_{bulk}^{L}}{\partial P_m^{L}} = 0, \qquad
		\frac{\partial f_{bulk}^{L}}{\partial P_s^{L}} = 0,
		\label{eq:Static_eqm}
	\end{equation} 
	
\begin{figure}[htbp]
		\centering
		\includegraphics[width=1\linewidth]{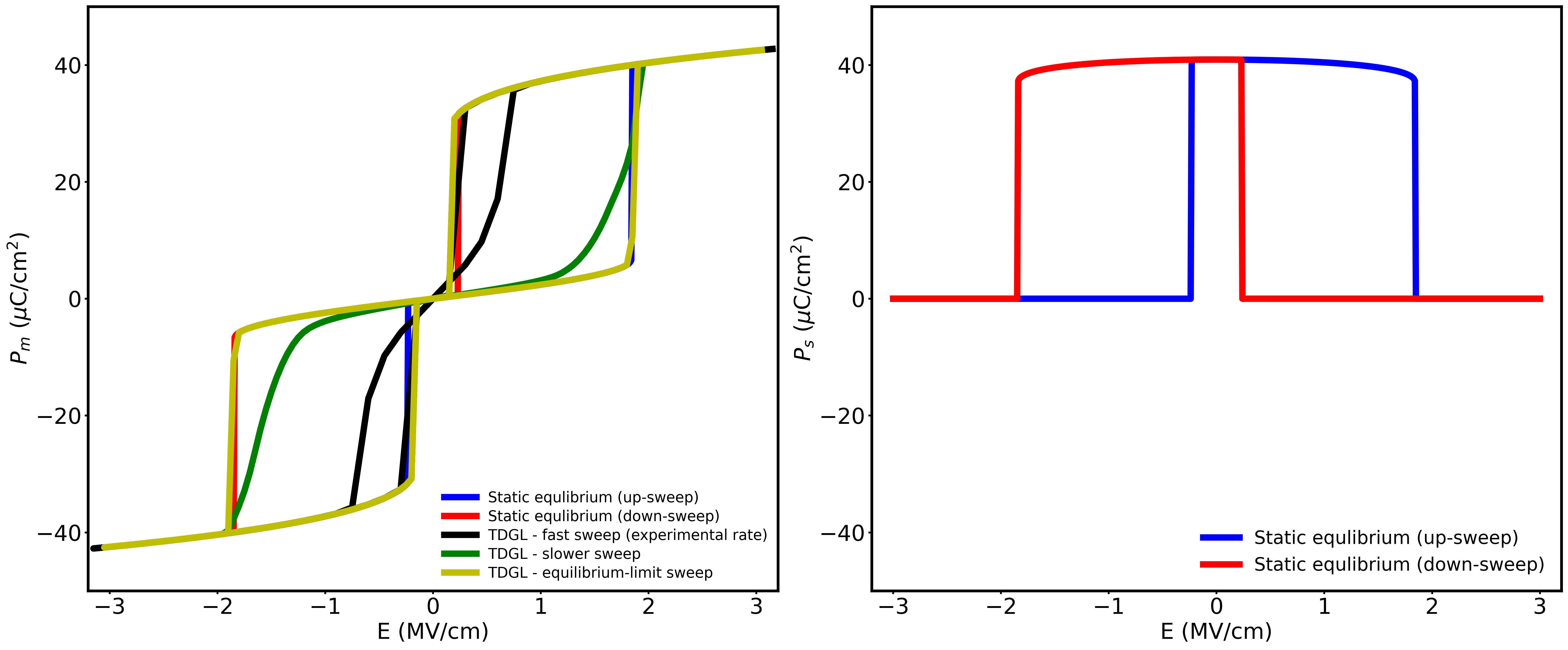}
		\caption{
			TDGL switching trajectories compared with the quasi-static equilibrium
			reference (red/blue). The experimental sweep rate ($\nu=10~\mathrm{kHz}$,
			black) exhibits kinetic deviation from equilibrium, while slower sweeping
			(green, yellow) progressively restores antipolar order and converges to the
			thermodynamic hysteresis loop.
		}
		\label{fig:sweep_rate_comparison}
	\end{figure}

	This procedure traces stable and metastable free-energy minima, and results in a sharp, discontinuous double hysteresis loop as shown in Fig.~\ref{fig:sweep_rate_comparison}(red-blue). At zero applied electric field, the system adopts an antipolar equilibrium state with $P_m \approx 0$ and two degenerate minima $P_s = \pm P_{s0} \approx \pm 40.9~\mu\mathrm{C/cm^2}$. As the field increases from zero, antipolar order destabilizes until a discontinuous AFE$\rightarrow$FE transition collapses $P_s$ to zero, producing a field-induced polar state with $P_{m0} \approx 40~\mu\mathrm{C/cm^2}$. Upon field reversal at $E=0$, restoration of antipolar order reproduces the double hysteresis loop. This behavior represents the ideal thermodynamic limit of AFE switching and is consistent with experimental observations in  epitaxial Pb-based AFE systems~\cite{Si_2025}.
	
	We next examine kinetic effects by evolving the same system using the TDGL equations and comparing the trajectories with the equilibrium reference. The macroscopic polarization is obtained by spatially averaging
	$P_{m,z}$ over the simulation domain~\cite{WANG2004749}. Fig.~\ref{fig:sweep_rate_comparison} shows the crossover from dynamic to
	quasi-static switching. At the experimental sweep rate ($\nu=10~\mathrm{kHz}$, black), the response deviates from equilibrium near $E\approx0$ because antipolar order does not fully recover, reflecting kinetic limitations rather than thermodynamic instability. Reducing the sweep rate progressively
	restores the equilibrium pathway: at intermediate rates (green), antipolar
	order fully re-emerges at zero field with $\pm P_s$ domains, while sufficiently slow sweeping
	(yellow) drives convergence to the quasi-static thermodynamic limit with a sharp symmetric $P$--$E$ loop. Continued coarsening in this single-crystal setting can bias the system toward a single-variant $P_s$ domain, reflecting kinetic selection in the absence of thermal fluctuations or microstructural disorder. Microstructures at characteristic field states for green-curve are shown in Fig.~S1.
	
	These results show that TDGL dynamics capture the crossover from kinetically limited switching at experimentally relevant sweep rates to the quasi-static thermodynamic limit. The experimentally calibrated field evolution is therefore retained to maintain realistic switching conditions, while the quasi-static electric-field loading serves only as a thermodynamic reference. 
	
	\subsection{Polycrystalline thin-film microstructures and architecture design}
	\label{sec:PolyX}
	
\begin{figure}[htbp]
		\centering
		\includegraphics[width=1.\textwidth]{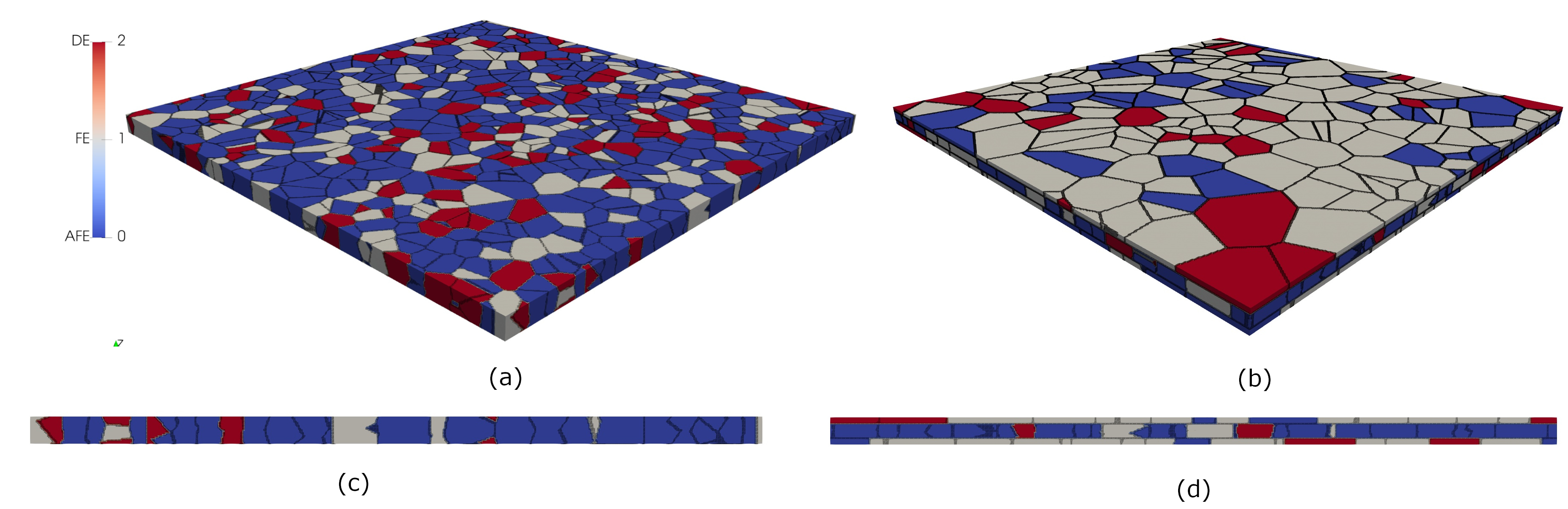}
		\label{fig:3L_micro}
		\caption{
			Representative polycrystalline thin-film microstructures used in the TDGL simulations.
			(a) Single-layer AFE-dominated film with minority FE and DE grains.
			(b) Multilayer film with vertically segregated phase distributions, consisting of
			AFE-rich and FE-rich layers.
			Panels (a,b) show oblique 3-D views, while panels (c,d) show the
			corresponding $y$–$z$ cross-sectional views. Grain boundaries are indicated in black.  The color scale denotes the phase-ID.
		}
		\label{fig:eg_microstructures}
	\end{figure}

	To examine the influence of microstructure on AFE--FE switching, the single-crystal
	framework is extended to polycrystalline thin films. 3-D polycrystalline geometries are generated using Voronoi tessellation, producing a grain-resolved topology with sharp, immobile grain boundaries. Grain assignment is based on spatial proximity in all three dimensions, yielding equiaxed rather than columnar grains. In multilayer architectures, grains are confined within individual layers and do not span layer interfaces. Material phase identity is assigned on a grain-by-grain basis, consistent with grain-scale control of phase stability in HZO systems~\cite{Park_2017}. Representative polycrystalline microstructures used in the TDGL simulations are shown in Fig.~\ref{fig:eg_microstructures}.

	The resulting grain-resolved structure introduces strong spatial heterogeneity in
	crystallographic orientation and material parameters. Each grain retains its
	assigned phase that determines its local Landau coefficients, dielectric response,
	and kinetic prefactor. While the polarization field remains continuous across grain
	boundaries, discontinuities in material parameters give rise to spatially varying
	switching barriers and coercive fields. Consequently, switching proceeds
	non-uniformly, with domain walls pinning at grain boundaries, deflecting along them,
	or propagating preferentially through clusters of more easily switchable grains.
	These effects emerge naturally from the TDGL formulation without invoking explicit
	grain-boundary mobility models.
	
	Grain sizes are chosen to reflect experimentally observed nanocrystalline regimes.
	GIXRD measurements on ultrathin ZrO$_2$ films report coherently diffracting domain
	sizes of approximately 5--15~nm for thicknesses between 5--20~nm, reflecting
	crystallite (coherence) lengths rather than physical grain diameters~\cite{JKD_ZrO2}.
	For polycrystalline Hf$_{1-x}$Zr$_x$O$_2$ thin films, experimental studies report
	laterally extended nanocrystalline grains with characteristic length scales on the
	order of $\sim$30~nm over a broad range of Zr compositions, spanning FE- and
	AFE-dominated electrical responses~\cite{SiJoonKim_Stress}.
	Based on these observations, we target a mean in-plane grain size of approximately
	30--35 grid points (30--35~nm for $l_0 = 1$~nm) by using about 800 grains in a
	$940\times940\times10$ grid. In multilayer films, grains are distributed among layers in proportion to layer thickness to maintain comparable mean grain sizes and layer-confined
	morphologies. Thicker layers therefore contain a larger number of grains without
	introducing structural bias.
	
	Throughout the TDGL simulations, grain morphology, phase assignment, phase fractions, simulation parameters, are prescribed a priori and remain unchanged. This reflects the experimental context of electric-field-driven switching in which grain structures are effectively immobile and microstructural evolution (e.g., grain growth or grain-boundary migration) occurs on much longer timescales. The analysis therefore focuses exclusively on polarization dynamics within a static polycrystalline network.
	
	\subsection{Grain-resolved switching microstructures in polycrystalline single-layer AFE films}
	\label{sec:1L_microstrucutres}
	
\begin{figure}[htbp]
		\centering
		\includegraphics[width=\linewidth]{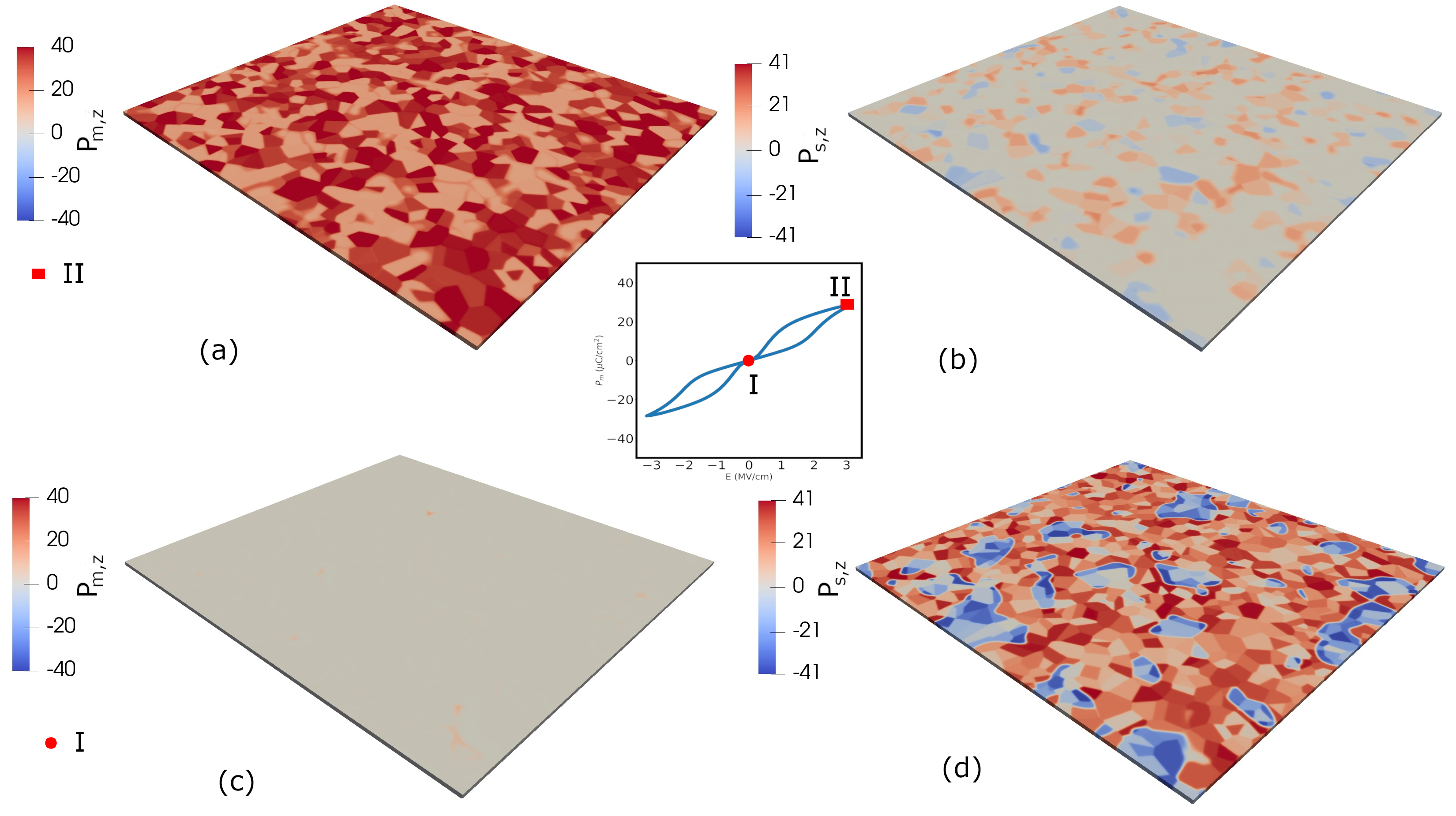}
		\caption{	
			Global-frame polarization microstructures in a 100\% AFE single-layer
			polycrystalline film.
			Panels (a,b) show $P_{m,z}^G$ and $P_{s,z}^G$ at $E=+E_{\max}$, and panels (c,d) show
			the corresponding fields at $E=0$ after field reversal. 
		}
		\label{fig:global_microstructures_AFE}
	\end{figure}

	For the calibrated AFE $P$--$E$ switching response shown in
	Fig.~\ref{fig:calib_AFE}, we examine the corresponding grain-resolved polarization
	microstructures. Fig.~\ref{fig:global_microstructures_AFE} shows such microstructures in global-frame during electric-field sweeping at two
	characteristic points along the $P$--$E$ curve: the maximum applied electric field
	($E = +E_{\max}$), corresponding to the field-induced polar state, and zero field
	after reversal ($E = 0$), corresponding to recovery of the antipolar ground state.
	The configuration at $E = -E_{\max}$ is symmetry-equivalent and is therefore not
	shown. Identical color scales (in $\mu$C/cm$^2$) are used for all polarization microstructure figures unless otherwise specified.
	
	Although all grains share identical AFE Landau coefficients, crystallographic
	misorientation introduces pronounced spatial heterogeneity, resulting in
	non-uniform switching through distributed nucleation, partial suppression of
	antipolar order under applied electric fields, and recovery of antipolar domains upon field
	reversal. Polarization microstructures of 100\% FE film
	(Fig.~\ref{fig:calib_FE}) are shown in Fig.~S2. In that
	case, $P_{s,z}$ remains identically zero throughout the field cycle.
	
	At $E = +E_{\max}$, $P_{m,z}^G$ approaches a
	near-saturated value $40 \mu$C/cm$^2$ [Fig.~\ref{fig:global_microstructures_AFE}(a)], indicating a field-stabilized
	FE-like state. $P_{s,z}^G$ is strongly reduced
	but remains finite
	[Fig.~\ref{fig:global_microstructures_AFE}(b)], despite the absence of any
	intrinsically FE phase. This residual antipolar signal reflects incomplete
	suppression of antipolar order arising from grain-scale heterogeneity rather than a
	uniform destabilization of the AFE state.
	
	Upon field reversal at $E = 0$, $P_{m,z}$ collapses nearly everywhere
	(Fig.~\ref{fig:global_microstructures_AFE}(c)), reflecting the absence of
	$P_r$ in a purely AFE system. Such reversible field-induced polar
	states have been observed experimentally in ultrathin ZrO$_2$ films, where
	confinement and processing conditions suppress the tetragonal-to-monoclinic
	transformation and stabilize a tetragonal-to-orthorhombic switching
	pathway~\cite{Muller_2012,Luo_2021}. In contrast, $P_{s,z}^G$ recovers strongly and
	forms extended antipolar domain patterns
	[Fig.~\ref{fig:global_microstructures_AFE}(d)], indicating restoration of the AFE
	ground state.

	Comparison with local-frame polarization fields (Fig.~S3) reveals that the residual $P_{s,z}^G$ at $E_{\max}$ arises from projection and incomplete cancellation of locally saturated antipolar domains across misoriented grains, rather than suppression of the intrinsic AFE order. These results demonstrate that strong electric fields drive a nominally single-phase AFE polycrystalline film into a reversible field-induced polar state, while
	antipolar order persists locally and fully recovers upon unloading. The macroscopic
	$P$--$E$ response therefore emerges from the collective behavior of
	crystallographically distinct grains, underscoring the importance of explicitly
	resolving microstructural effects when interpreting experimental switching
	characteristics. Similar grain-mediated disruption and recovery of antipolar order
	have been exploited in Pb-based perovskite AFEs to tailor hysteresis and energy
	storage performance~\cite{Zhou_2025,Yang_2025}.

	\subsection{Influence of phase fractions on macroscopic switching response}
	\label{subsec:phase_fraction_compare}

\begin{figure}[htbp]
		\centering
		\includegraphics[width=0.5\linewidth]{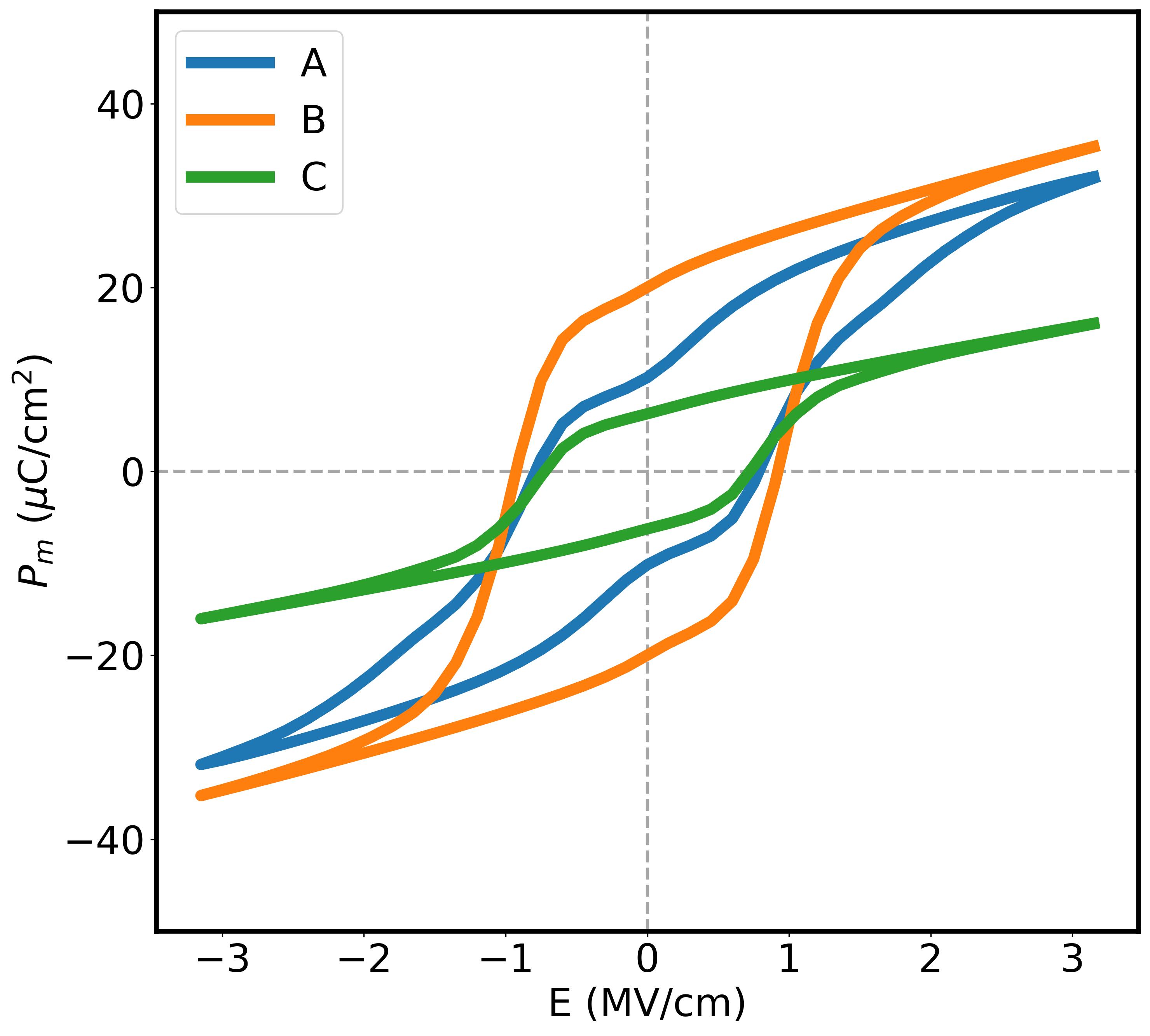}
		\caption{
			Simulated $P$--$E$ loops for single-layer polycrystalline films using
			experimental-informed phase fractions from Ref.~\cite{Mittmann}:
			(A) 60\% AFE–35\% FE–5\% DE,
			(B) 15\% AFE–75\% FE–10\% DE, and
			(C) 5\% AFE–25\% FE–70\% DE.
		}
		\label{fig:PE_phase_fraction_compare}
	\end{figure}

To connect the phase-field simulations with experimentally observed
$P$--$E$ trends, we adopt representative phase fractions reported for
HZO thin films by Mittmann \emph{et al.}~\cite{Mittmann}. These
experimentally inferred fractions are: (A) 60\% AFE–35\% FE–5\% DE,
(B) 15\% AFE–75\% FE–10\% DE, and
(C) 5\% AFE–25\% FE–70\% DE. In the present simulations, these phase
fractions are treated as prescribed input parameters, while intrinsic
phase thermodynamics are represented by fixed Landau coefficients
(see Section~\ref{subsec:parameter_calibration}). This approach isolates
microstructure- and architecture-driven switching effects. For each case,
phase identities are randomly assigned to grains according to the
prescribed fractions. The resulting simulated $P$–$E$ loops are shown in
Fig.~\ref{fig:PE_phase_fraction_compare}.

	When the AFE phase dominates (A), the $P$--$E$ loop remains pinched at
	$E=0$ with low $P_r$, indicating some recovery of $P_s$
	upon field reversal, in line with experimentally inferred thermodynamic
	scenarios for pinched hysteresis in ZrO$_2$- and HfO$_2$-based films~\cite{Lomenzo2020}.
	As the FE fraction increases (B), the response
	evolves toward a FE-like hysteresis with finite remanence, although $P_r$ remains
	reduced relative to 100\% FE film due to AFE and DE phases.
	When the DE phase dominates (C), both remanent and saturation polarizations are
	further suppressed, reflecting dilution of switchable polar regions and an
	increased contribution from linear dielectric response.
	
	%when the external electric field is reduced back to zero, the system does not stay in the field-induced polar state, but instead relaxes back to a state with little or no net polarization because that state is energetically favored at low field.
	
	Overall, the simulated $P$--$E$ curves show similar trends as reported by Mittmann \emph{et al.}~\cite{Mittmann}, demonstrating
	that the macroscopic switching response is governed primarily by relative phase
	fractions rather than by specific chemical mechanisms or additional effects not
	explicitly included in the model.

	\subsection{Vertical phase architectures in polycrystalline thin films}
	
	We next examine whether vertical phase arrangement provides an additional degree of control over $E_c$ and $P_r$. To this end, we systematically compare single-layer and multilayer architectures under otherwise identical conditions, isolating the effect of vertical phase segregation from compositional influences.
	
	Single-layer films consist of mixed AFE and FE grains distributed throughout the full film thickness, producing fully 3-D equiaxed grains with grain boundaries extending in all spatial directions. Electrical coupling to the electrodes therefore occurs through percolative pathways within a heterogeneous grain network, and these films serve as the reference.
	
\begin{figure}[htbp]
		\centering
		\includegraphics[width=0.95\linewidth]{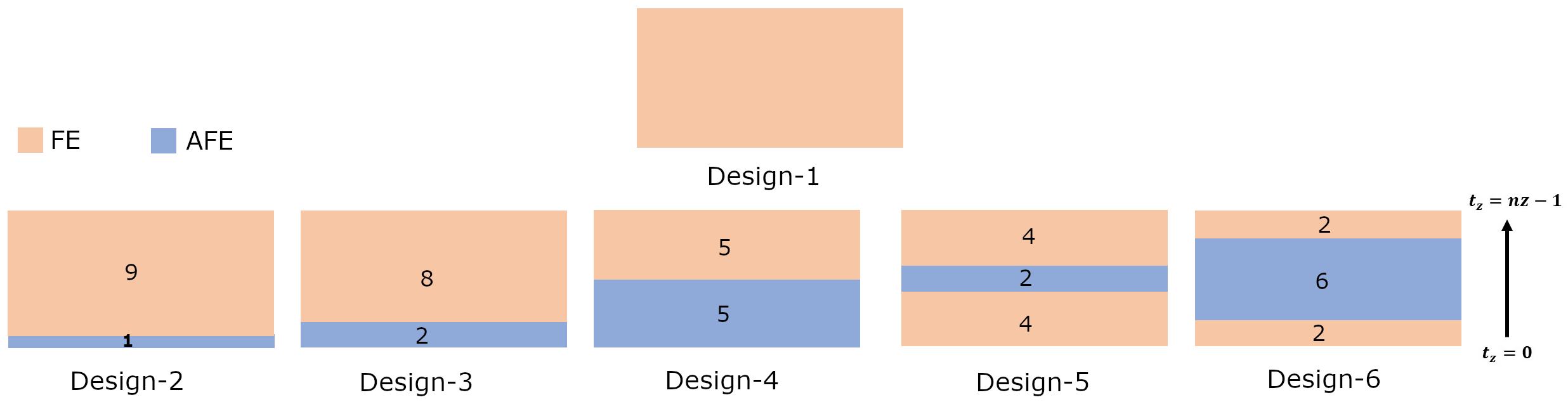}
		\caption{
Schematic illustration of the polycrystalline thin-film architectures
considered in this study: Design-1 shows the single-layer configuration,
Designs-2--4 correspond to double-layer architectures, and Designs-5--6
represent triple-layer architectures. Numbers indicate relative layer
thicknesses (grid units) along the film-normal direction.
		}
		\label{fig:design_space}
	\end{figure}

	Multilayer films are constructed by vertically segregating phases along the film normal. Two classes are considered. In double-layer architectures, an AFE-rich layer contacts the bottom electrode and is capped by an FE-rich layer, resulting in asymmetric electrode--phase contact. In triple-layer architectures, the AFE-rich layer is confined to the film interior and sandwiched between FE-rich layers that contact both electrodes, preserving symmetric electrode contact while isolating the AFE phase from the interfaces. A schematic overview of the architectures is shown in Fig.~\ref{fig:design_space}, and a representative polycrystalline film architecture is shown in Fig.~\ref{fig:eg_microstructures}.
	
	The total film thickness, phase identity, and grain-size statistics are kept identical across all architectures. In multilayer films, the placement of an AFE-rich region is systematically varied relative to the electrodes without altering the intrinsic properties of the FE-dominated material. This controlled design space enables direct evaluation of the influence of vertical phase distribution on $P_r$ and $E_c$. Internal FE--AFE interfaces influence switching through local electrostatic and kinetic coupling, represented here in a minimal phenomenological manner, such that observed trends reflect geometry- and coupling-driven effects rather than changes in intrinsic material composition. Phase fractions for all architectures are summarized in Table~\ref{tab:phase_fractions}.

	\begin{table}[H]
		\centering
		\caption{Phase fractions used in the polycrystalline thin-film architectures.}
		\label{tab:phase_fractions}
		\begin{tabular}{lcc}
			\hline
			Design type & AFE fraction (\%) & FE fraction (\%) \\
			\hline
			Single-layer (Baseline~1) & 10 & 90 \\
			Single-layer (Baseline~2) & 30 & 70 \\
			FE-rich layer (multilayer) & 10 or 30 & 90 or 70 \\
			AFE-rich layer (multilayer) & 60 & 40 \\
			\hline
		\end{tabular}
	\end{table}

	\subsubsection{Macroscopic switching response of multilayer architectures}
	
\begin{figure}[t]
		\centering
		\begin{subfigure}[b]{0.48\linewidth}
			\centering
			\includegraphics[width=\linewidth]{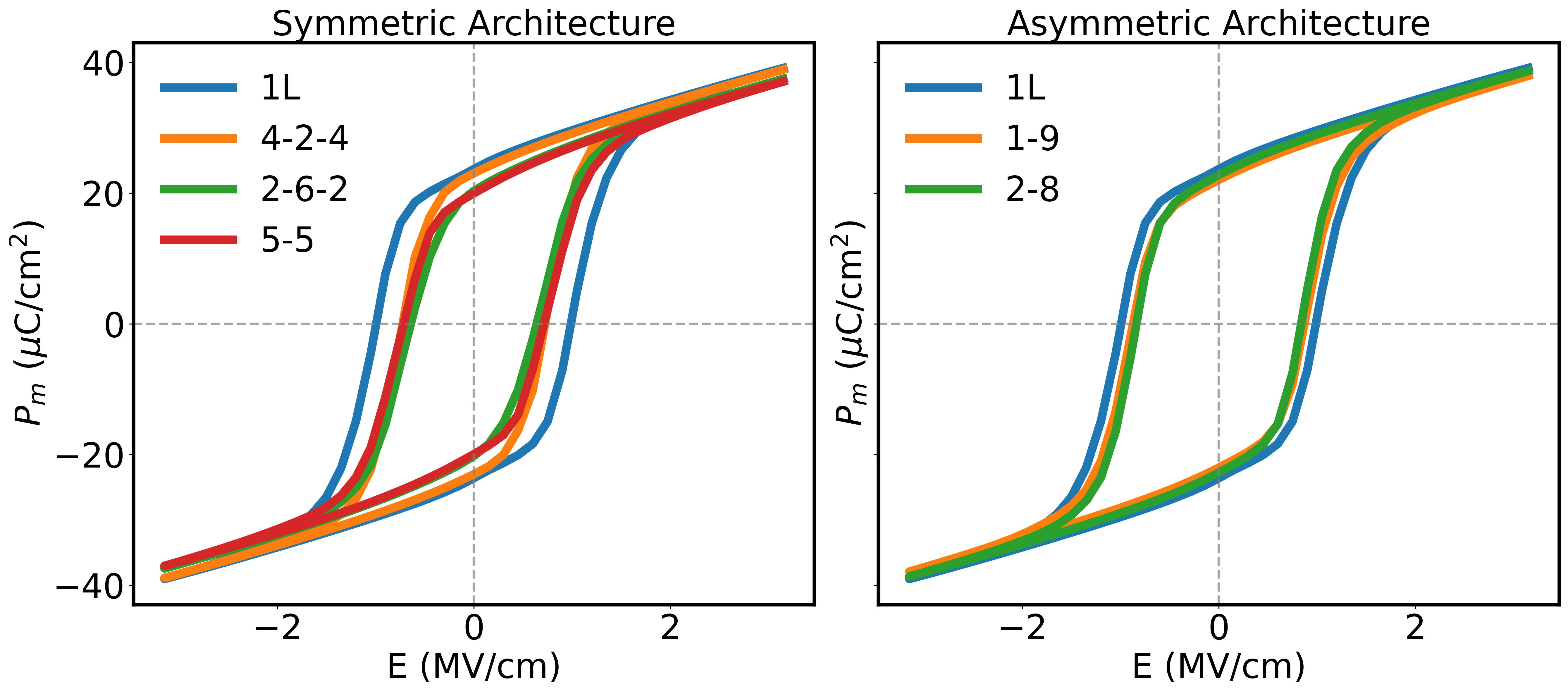}
			\caption{Baseline~1: AFE 10\%--FE 90\%}
			\label{fig:poly_PE_10AFE}
		\end{subfigure}
		\hfill
		\begin{subfigure}[b]{0.48\linewidth}
			\centering
			\includegraphics[width=\linewidth]{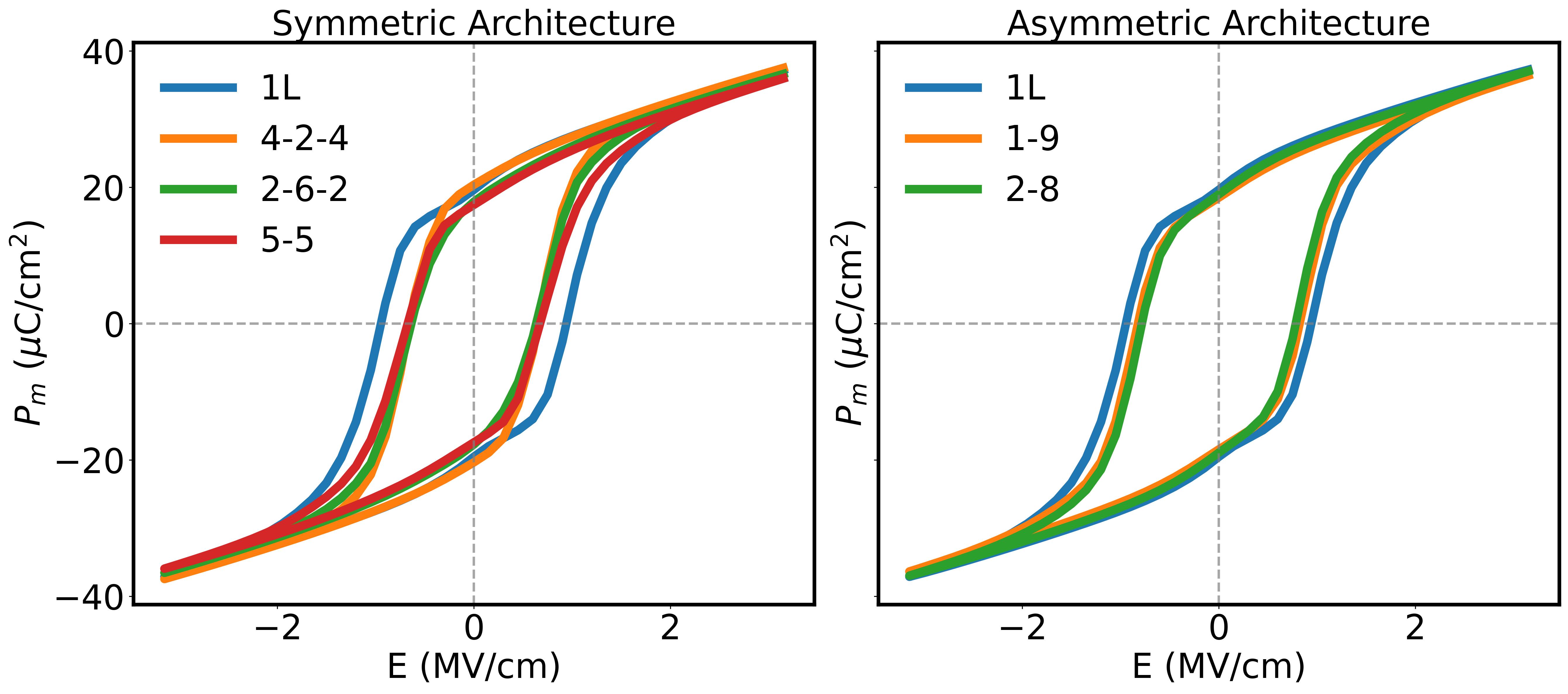}
			\caption{Baseline~2: AFE 30\%--FE 70\%}
			\label{fig:poly_PE_30AFE}
		\end{subfigure}
		\caption{$P$--$E$ response of polycrystalline thin films for the single-layer and
			multilayer architectures. (a) Architectures with FE-rich layers having
			Baseline~1 phase fractions. (b) Architectures with FE-rich layers having
			Baseline~2 phase fractions.}
		\label{fig:poly_PE_loops}
	\end{figure}

Fig.~\ref{fig:poly_PE_loops} compares the $P$–$E$ response of the two
single-layer baseline films and the different multilayer architectures.
These $P$--$E$ curves correspond to a single grain-topology
		realization; statistical variability across multiple realizations is
		quantified in the next subsection.
As shown in Section~\ref{subsec:phase_fraction_compare}, both single-layer
films exhibit predominantly FE-like single hysteresis loops despite the
presence of a finite AFE fraction. Introducing vertical phase segregation
through multilayer architectures systematically modifies the
$P$--$E$ response. In particular, films containing an AFE-dominated layer
exhibit reduced loop area relative to the baseline films, motivating a
quantitative comparison of $P_r$ and $E_c$. Notably, all multilayer
architectures retain a single, continuous hysteresis loop without typical
AFE double hysteresis. This indicates that the AFE-dominated layer does
not switch independently but instead modulates the macroscopic response
through electrostatic and kinetic coupling to surrounding FE-rich regions.
Such behavior is seen in prior phase-field studies, which report
preferential FE domain nucleation at AFE domain boundaries and suppression
of saturation polarization through interfacial coupling in AFE-containing
structures~\cite{Zhu_2023,Xu_2025}. Vertical phase placement therefore
manifests primarily as systematic shifts in $P_r$ and $E_c$, while
preserving an overall FE-like switching character. A stronger AFE-rich
layer (e.g., 90\% AFE) would further reduce $E_c$ while also suppressing
$P_r$, leading to poor polarization retention at $E=0$. Such sensitivity
of AFE switching pathways to electrostatic boundary conditions and layer
stacking has been demonstrated experimentally in dielectric/AFE
heterostructures that stabilize otherwise unstable polarization states in
zirconia-based films~\cite{Hoffmann2022}.

	\subsubsection{Quantitative analysis of remanent polarization and coercive field}
	
	To quantitatively assess the influence of vertical phase distribution on switching
	behavior, we extract $P_r$ and $E_c$ from the $P$--$E$ loops shown in
	Fig.~\ref{fig:poly_PE_loops}. For each architecture,
	$P_r$ and $E_c$ are evaluated separately for positive and negative field sweeps and
	then averaged to obtain representative values,
	\begin{equation}
		P_{r,\mathrm{avg}} = \tfrac{1}{2}\left(|P_r^{+}| + |P_r^{-}|\right), \qquad
		E_{c,\mathrm{avg}} = \tfrac{1}{2}\left(|E_c^{+}| + |E_c^{-}|\right),
	\end{equation}
	where superscripts $+$ and $-$ denote quantities extracted from the positive and
	negative field branches, respectively. These averaged metrics provide a
	representative measure of intrinsic switching behavior while minimizing the
	influence of minor loop asymmetries.
	
\begin{figure}[t]
		\centering
		\includegraphics[width=1.\textwidth]{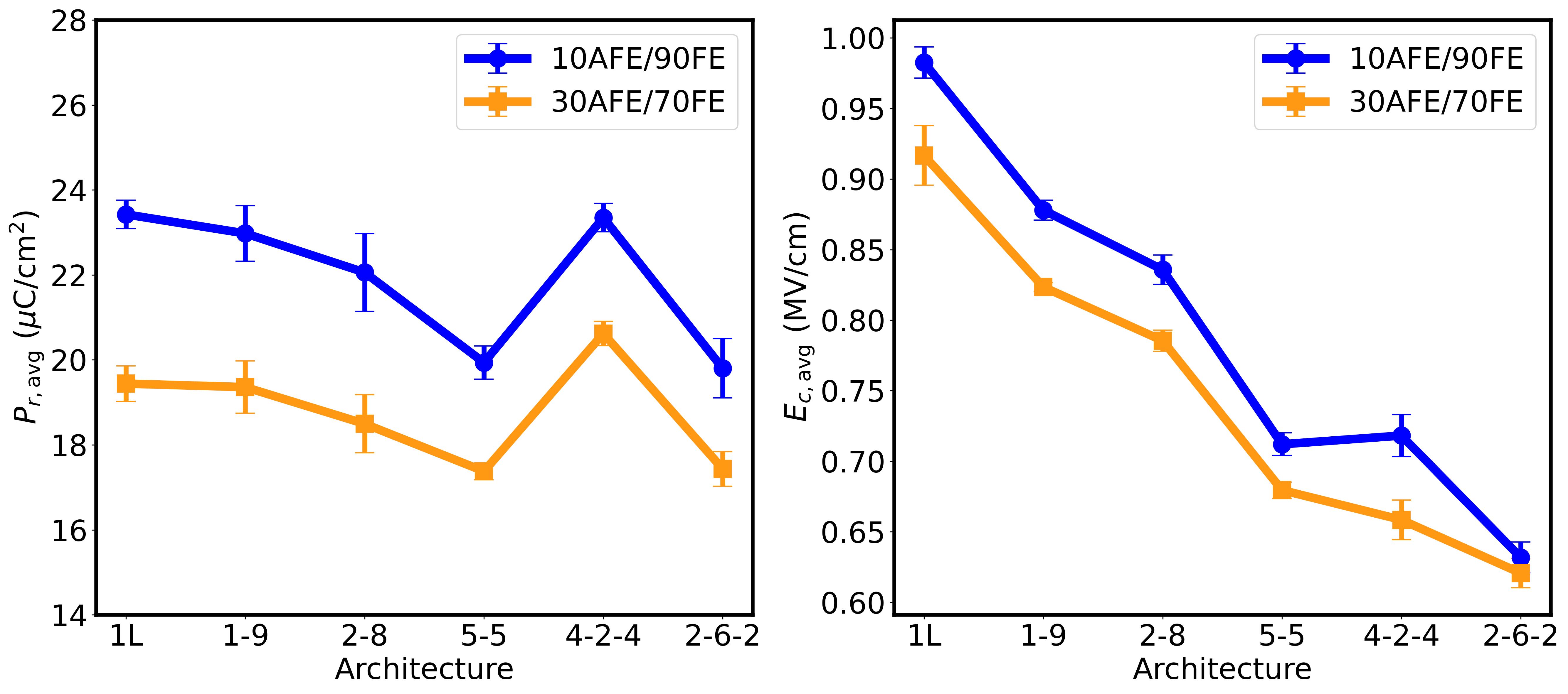}
		\caption{Average remanent polarization $P_{r,\mathrm{avg}}$ and coercive field
			$E_{c,\mathrm{avg}}$ for polycrystalline thin films with different vertical
			phase distributions. Symbols denote mean values obtained from multiple grain-topology realizations, and error bars indicate one standard deviation reflecting microstructural.}
		\label{fig:PrEc_metrics}
	\end{figure}

	Fig.~\ref{fig:PrEc_metrics} summarizes $P_{r,\mathrm{avg}}$ and
	$E_{c,\mathrm{avg}}$ for all polycrystalline architectures.
	Each data point represents the mean value obtained from multiple
			independent grain-topology realizations, and the error bars denote the
			corresponding standard deviation, thereby quantifying the variability arising
			from stochastic Voronoi microstructures.
	For both baseline fractions, the single-layer films exhibit the largest $P_{r,\mathrm{avg}}$ and $E_{c,\mathrm{avg}}$.
	Introducing an AFE-dominated layer in multilayer architectures leads to a
	systematic reduction in both quantities. Notably, the reduction in $E_{c,\mathrm{avg}}$ is
	consistently more pronounced than that in $P_{r,\mathrm{avg}}$, indicating that the primary role
	of the AFE-rich layer is to lower the effective switching field rather than to
	strongly modify $P_{r,\mathrm{avg}}$.  The numerical values corresponding to Fig.~\ref{fig:PrEc_metrics} are provided in Tables~S1 and S2.
	
	Experimental studies on FE/AFE sandwich-stacked HfO$_2$-based thin films have
	similarly reported substantial reductions in $E_c$ in multilayer architectures,
	with some stacks also exhibiting enhanced $P_r$~\cite{Sandwich_HfO2_Study}. In the present simulations, vertical phase arrangement robustly reduces $E_c$ through electrostatic and kinetic
	coupling between FE-rich and AFE-rich layers, but does not generically enhance
	$P_r$. This reflects the absence of interface- or chemistry-driven AFE$\rightarrow$FE phase conversion during electric-field cycling. Consequently,
	$P_r$ is governed primarily by the volume fraction and connectivity of FE-rich
	layers, whereas vertical stacking mainly reshapes switching pathways rather than
	the zero-field ground state. Additional interface-, dopant-, or processing-induced phase stabilization mechanisms are therefore likely required to achieve substantial $P_r$ enhancement beyond the architecture-only effects captured here; explicit
	incorporation of defect-mediated electrostatic and transport effects will
	be addressed in future work.

Among symmetric multilayer architectures, increasing the thickness of the
AFE-dominated layer reduces both $P_{r,\mathrm{avg}}$ and
$E_{c,\mathrm{avg}}$, reflecting the growing
volume fraction of AFE-rich material that resists ferroelectric switching
and redistributes the applied electric field. In contrast,
asymmetric multilayer architectures partially preserve $P_{r,\mathrm{avg}}$ while still
achieving a substantial reduction in$E_{c,\mathrm{avg}}$, yielding a more favorable
trade-off between polarization retention and switching field.
Across all multilayer configurations, the reduction in
		$E_{c,\mathrm{avg}}$ persists despite moderate statistical variability,
		suggesting that the trend is linked to the imposed vertical phase
		arrangement rather than stochastic grain connectivity.
In the following, we analyze the grain-scale switching pathways responsible
for these trends, before distilling general design principles.

	\subsubsection{Grain-resolved switching mechanisms and design implications}
	\label {grain_resolved_path}
	
	To identify the microscopic origin of the reduced $E_c$, we analyze grain-resolved switching pathways for architectures having Baseline~1 fractions, focusing on a symmetric 2--6--2 (largest $E_c$ reduction) and an asymmetric 1--9 (strongest vertical asymmetry). Baseline~2 architectures show similar behavior and are omitted for brevity. Because $E_c$ differs between architectures, snapshots are aligned by switching events rather than applied-field values, enabling direct comparison of nucleation, percolation, and system-spanning reversal.
	
\begin{figure}[htbp]
		\centering
		
		% ---------- TOP BLOCK ----------
		\begin{minipage}[t]{0.7\linewidth}
			\centering
			
			% (a)
			\includegraphics[width=\linewidth]{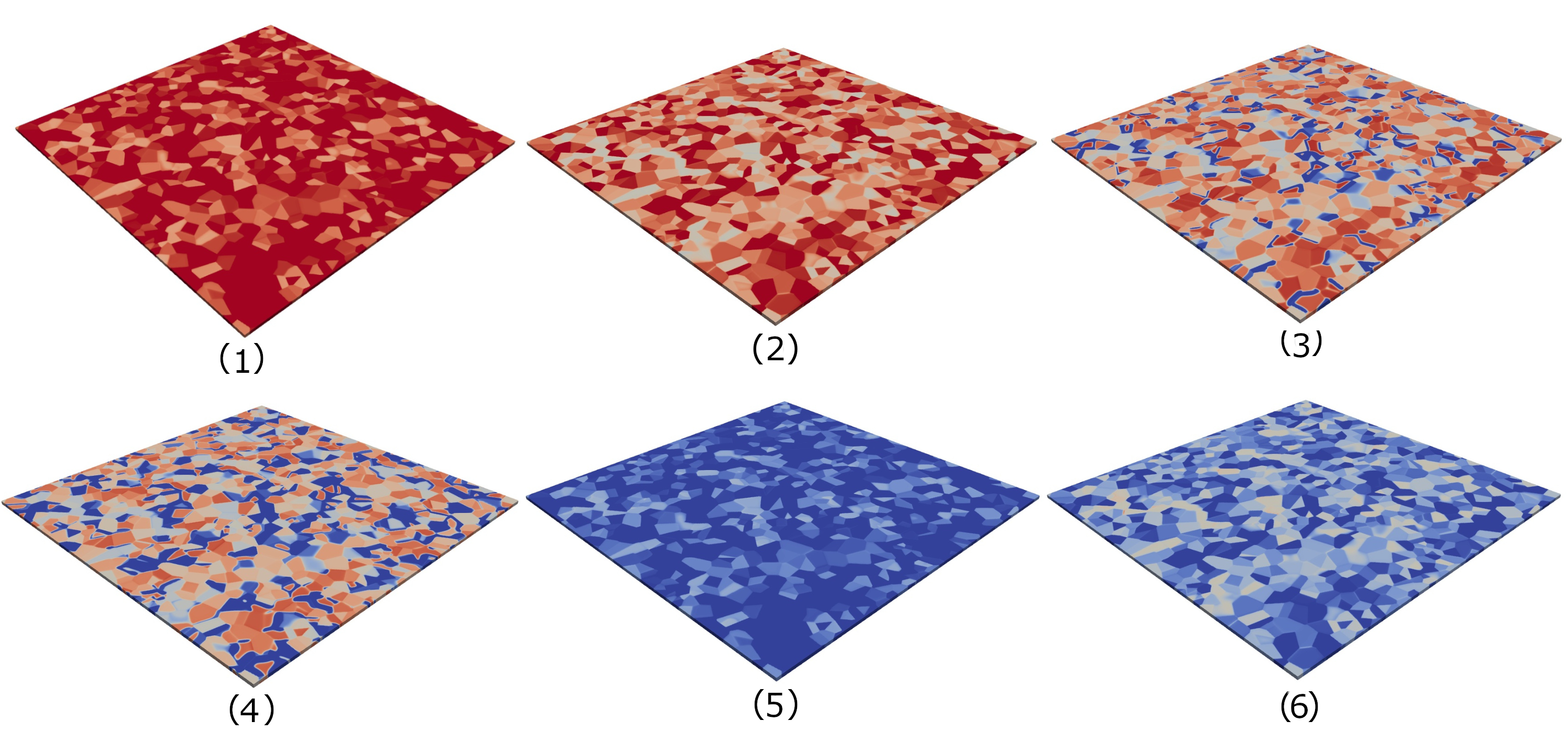}\\[-2pt]
			{\small (a) Single-layer (Baseline~1)}\\[8pt]
			
			% (b)
			\includegraphics[width=\linewidth]{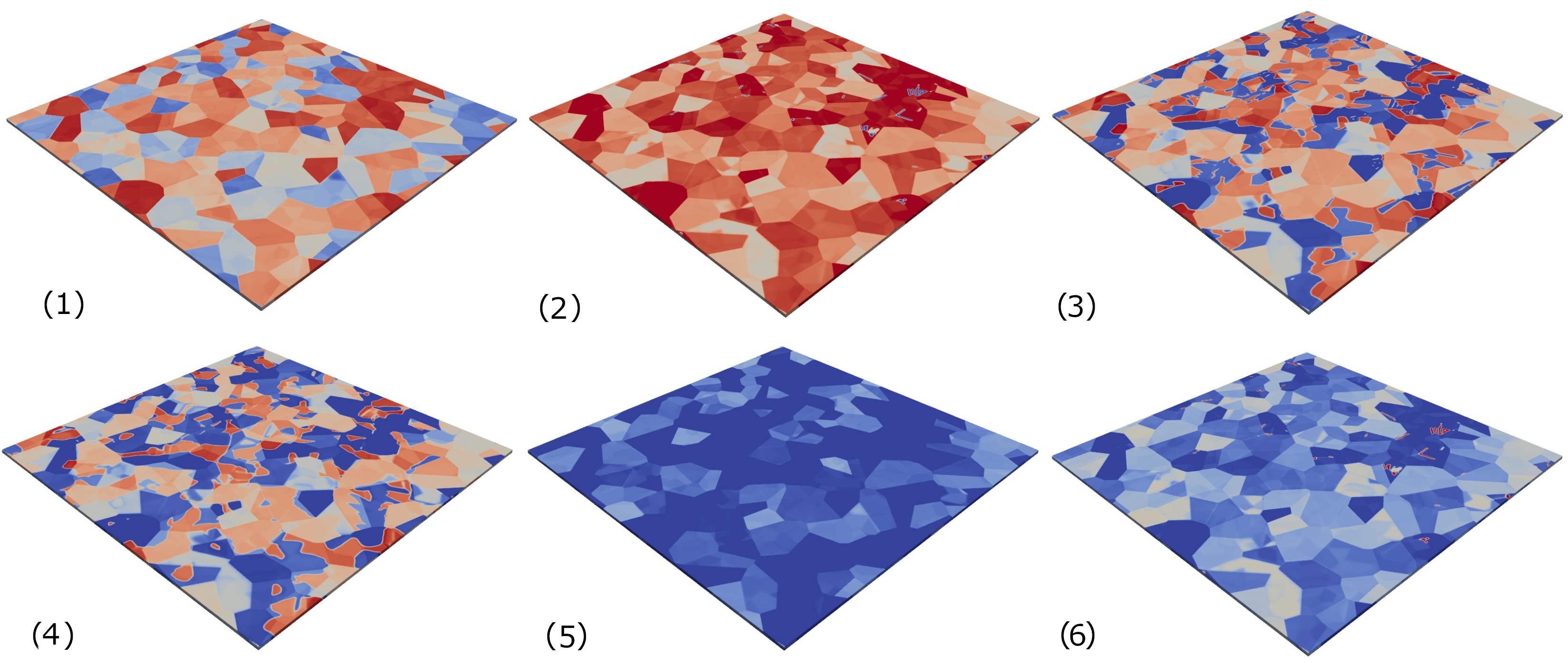}\\[-2pt]
			{\small (b) Symmetric multilayer (2--6--2)}\\[8pt]
			
			% (c)
			\includegraphics[width=\linewidth]{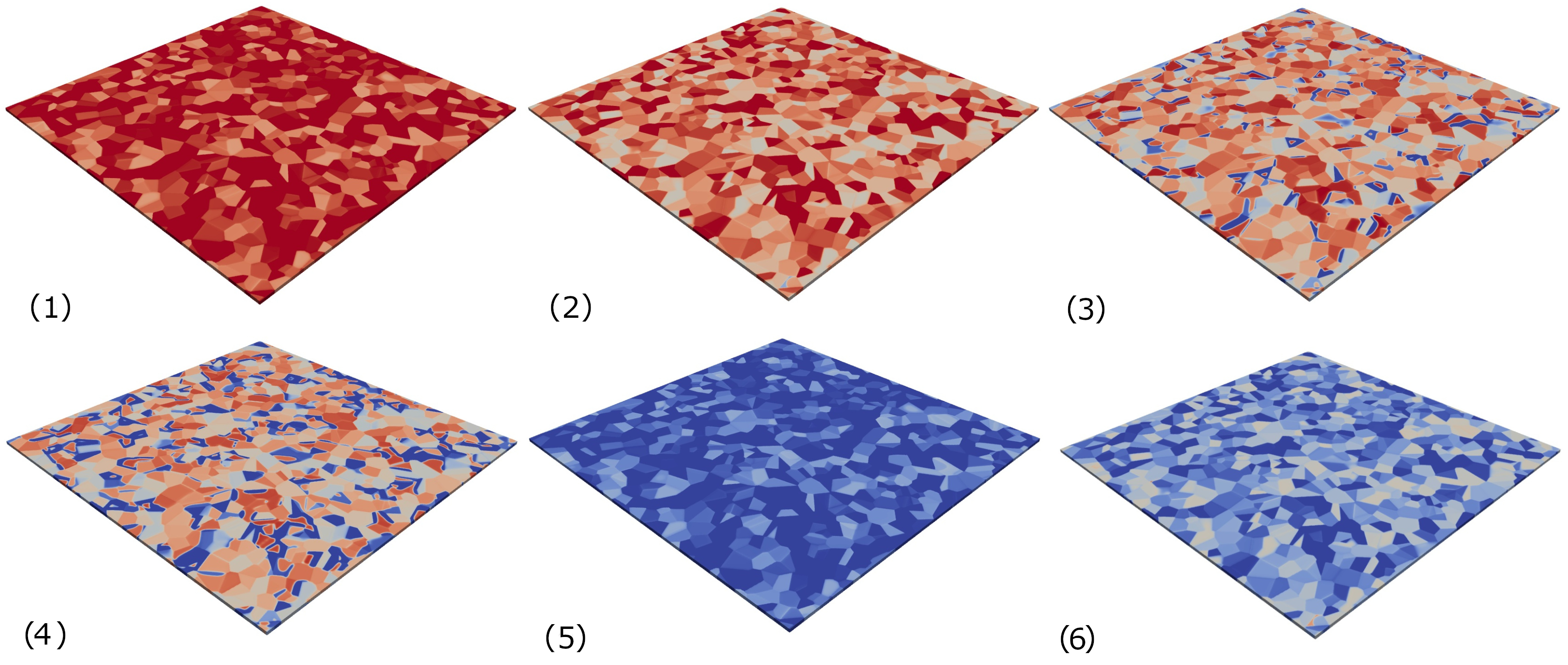}\\[-2pt]
			{\small (c) Asymmetric multilayer (1--9)}
			
		\end{minipage}
		\hfill
		\begin{minipage}[c]{0.25\linewidth}
			\centering
			\vspace*{0.18\textheight} % tune if needed
			
			% (d)
			\includegraphics[width=\linewidth]{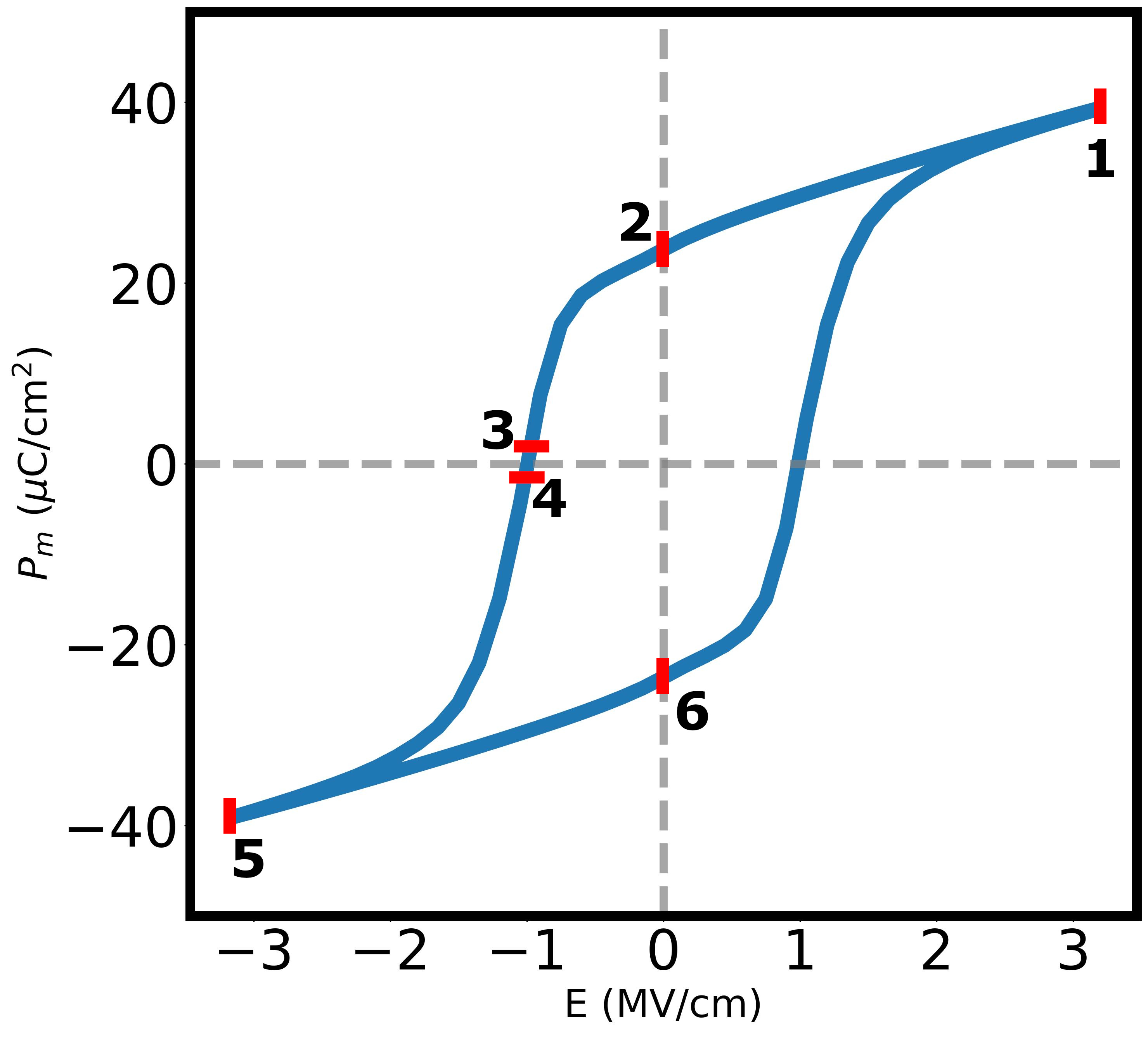}
			{\small (d) $P$--$E$ curve}\\[4pt]
		\end{minipage}
		
		\vspace{8pt}
		
		% ---------- COLORBAR ----------
		\begin{minipage}{0.7\linewidth}
			\centering
			
			% (e)
			\includegraphics[width=\linewidth]{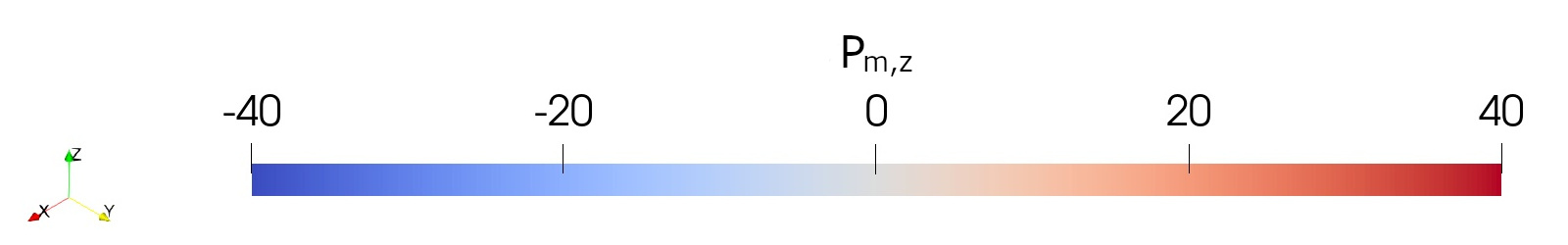}
		\end{minipage}
		
		\caption{Grain-resolved switching snapshots of $P_{m,z}$. Panels (1--6) in each sub-figure correspond to equivalent, event-aligned stages marked on the $P$--$E$ curve.}
		\label{fig:grain_snapshots}
	\end{figure}

	Six snapshots are extracted at equivalent stages of the hysteresis cycle (Fig.~\ref{fig:grain_snapshots}): (a) $+E_{\max}$ saturation, (b) positive remanence at $E=0$, (c,d) the coercive window bracketing macroscopic reversal, (e) $-E_{\max}$ saturation, and (f) negative remanence at $E=0$. The coercive window shifts from $E=-0.90/-1.05$~MV/cm in the single-layer film (Fig.~\ref{fig:grain_snapshots}(a)-3,4), to $E=-0.75/-0.90$~MV/cm in the asymmetric multilayer (Fig.~\ref{fig:grain_snapshots}(c)-3,4), and to $E=-0.60/-0.75$~MV/cm in the symmetric multilayer (Fig.~\ref{fig:grain_snapshots}(b)-3,4), reproducing the hierarchy of $E_c$ values in Fig.~\ref{fig:PrEc_metrics}a.
	
	Such separation between single-domain and multidomain switching fields has also
	been reported in atomic-scale simulations of AFE PbZrO$_3$, highlighting the role
	of domain structures and electrostatic boundary conditions in lowering the
	effective switching field~\cite{PbZrO3_MLMD_Yubai,Dawber_2003}. In both those
	studies, switching proceeds via local collapse and
	reconfiguration of antipolar order, with strongly pinned domain walls, rather
	than through uniform domain-wall glide. The resulting reduction in $E_c$ thus
	originates from earlier nucleation and faster percolation of reversed regions. Such domain-structure–assisted switching is a generic consequence of polarization
	instabilities in ferroic systems, where
	electrostatic and gradient energies favor multidomain pathways~\cite{Bishop_2024}.
	
	In the single-layer Baseline~1 film, polarization
	reversal initiates through spatially isolated switched clusters at
	$E=-0.90$~MV/cm (Fig.~\ref{fig:grain_snapshots}(a)-3), with a system-spanning reversed
	network forming only at $E=-1.05$~MV/cm (Fig.~\ref{fig:grain_snapshots}(a)-4), resulting in
	the largest $E_c$. Such
	nucleation-and-growth–dominated switching mediated by grain boundaries is
	characteristic of FE polycrystals~\cite{CHOUDHURY20055313}. By contrast, the symmetric 2--6--2 architecture exhibits
	earlier nucleation at $E=-0.60$~MV/cm (Fig.~\ref{fig:grain_snapshots}(b)-3), followed by
	rapid formation of connected reversed pathways at a slightly higher field
	(Fig.~\ref{fig:grain_snapshots}(b)-4), yielding the lowest $E_c$ among all configurations.
	The asymmetric 1--9 architecture shows intermediate behavior, with earlier
	nucleation than the single-layer film at $E=-0.75$~MV/cm
	(Fig.~\ref{fig:grain_snapshots}(c)-4), but less uniform connectivity, leading to complete
	reversal only at $E=-0.90$~MV/cm (Fig.~\ref{fig:grain_snapshots}(c)-4). This reflects
	orientation-dependent switching heterogeneity induced by the asymmetric placement
	of the AFE-rich layer~\cite{CHOUDHURY20071415}.

	The spatial progression of switching can be quantified by the reversed-volume
	fraction $f_{\mathrm{rev}} = V(P_{m,z}<0)/V_{\mathrm{tot}}$.
	Table~\ref{tab:frev_snapshots} shows that the increase in $f_{\mathrm{rev}}$
	across the coercive window is largest for the symmetric multilayer, followed by
	the asymmetric multilayer and the single-layer film. The larger reversed fractions
	at comparable switching stages confirm that macroscopic reversal occurs at a
	lower percolation threshold in multilayer architectures.
	
	\begin{table}[h]
		\centering
		\caption{Reversed-volume fraction $f_{\mathrm{rev}}$ evaluated at equivalent,
			event-aligned stages of the hysteresis cycle for different film architectures.}
		\label{tab:frev_snapshots}
		\begin{tabular}{lcccccc}
			\hline
			Design 
			& $+E_{\max}$ 
			& $E=0,+\mathrm{rem}$ 
			& pre-rev 
			& post-rev 
			& $-E_{\max}$ 
			& $E=0,-\mathrm{rem}$ \\
			& $f_{\mathrm{rev}}$ 
			& $f_{\mathrm{rev}}$ 
			& $f_{\mathrm{rev}}$ 
			& $f_{\mathrm{rev}}$ 
			& $f_{\mathrm{rev}}$ 
			& $f_{\mathrm{rev}}$ \\
			\hline
			1L      & 0.00 & 1.58 & 28.21 & 45.41 & 100.00 & 97.24 \\
			2--6--2 & 0.00 & 0.50 & 38.47 & 56.66 & 100.00 & 98.92 \\
			1--9    & 0.00 & 1.27 & 25.06 & 43.00 & 100.00 & 98.11 \\
			\hline
		\end{tabular}
	\end{table}
	
	Across all architectures, $E_c$ is governed by the field required to establish a connected, system-spanning reversed network rather than by isolated local nucleation events, as reported in prior grain-resolved phase-field studies of ferroelectric polycrystals~\cite{CHOUDHURY20055313,CHOUDHURY20071415}. Multilayer architectures reduce $E_c$ by enabling earlier nucleation and more rapid percolation of reversed regions, reflecting the general tendency of multidomain switching pathways to lower effective switching fields in antiferroelectrics~\cite{PbZrO3_MLMD_Yubai}. Remanent snapshots in Fig.~\ref{fig:grain_snapshots} further reveal increased polarization heterogeneity at zero field in multilayer films. For Baseline~1, this heterogeneity suppresses polarization retention, whereas for Baseline~2, selected multilayers—most notably the symmetric 4--2--4—slightly exceed the single-layer $P_r$ while still exhibiting a reduced $E_c$ (Fig.~\ref{fig:PrEc_metrics}).
	
	\begin{center}
		\fbox{
			\begin{minipage}{0.95\linewidth}
				\textbf{Design principles for FE--AFE polycrystalline heterostructures
					(Fig.~\ref{fig:design_space})}
				
				\begin{itemize}
					\item \textbf{Vertical phase segregation reduces $E_c$.}
					Introducing an AFE-dominated layer within an FE-rich polycrystalline film lowers $E_c$ through electrostatic and kinetic coupling, without activating independent AFE switching.
					
					\item \textbf{Interior placement of the AFE layer is most effective.}
					Symmetric multilayer architectures with the AFE-rich layer confined to the film interior facilitate earlier percolation of reversed domains and yield the largest reduction in $E_c$, whereas asymmetric architectures provide a more moderate reduction.
					
					\item \textbf{$P_r$ is set by FE-rich layers.}
					The remanent polarization $P_r$ is governed primarily by the phase fraction and connectivity of FE-rich layers and cannot be enhanced by  vertical arrangement alone.
					
					\item \textbf{Optimal performance requires a balanced AFE fraction.}
					Moderately AFE-rich interior layers (e.g., 60\% AFE) provide a favorable trade-off
					between reduced $E_c$ and retained $P_r$, while higher AFE fractions further lower
					$E_c$ at the expense of polarization retention.
				\end{itemize}
				
		\end{minipage}}
	\end{center}
	
	\section{Conclusion}
	We developed a unified grain-resolved, 3-D phase-field framework to
	investigate electric-field-driven switching in polycrystalline hafnia-based thin
	films with mixed FE, AFE, and DE phases. The model explicitly resolves realistic 3-D grain topology, crystallographic orientation, and antipolar sublattice kinetics by evolving both
	macroscopic and staggered polarization order parameters, while adopting a uniaxial
	capacitor formulation with self-consistent depolarization fields. All thermodynamic
	and kinetic parameters were calibrated to experimental $P$--$E$ data and held fixed,
	enabling direct isolation of microstructural and architectural effects.
	
	The simulations show that polycrystallinity fundamentally reshapes AFE switching
	by promoting domain-structure--assisted pathways that reduce $E_c$ relative to single-crystal thermodynamic limits. In nominally
	single-phase AFE polycrystals, strong fields stabilize a reversible FE-like state
	while antipolar order persists locally due to crystallographic misorientation,
	showing AFE $\rightarrow$ FE transition. For mixed-phase single-layer films, the variation of $P_r$ and $E_c$ with phase fraction broadly follows experimentally observed trends, suggesting that the macroscopic switching response is governed primarily by phase distribution rather than by specific chemical stabilization mechanisms.

	Vertical phase arrangement provides an additional and effective control knob to tailor $P_r$ and $E_c$.
	Introducing an AFE-dominated layer within an FE-rich polycrystalline film reduces
	$E_c$ more strongly than $P_r$, while preserving
	a single, continuous FE-like hysteresis loop and avoiding independent AFE
	switching. Grain-resolved analysis reveals that this reduction in $E_c$ originates
	from earlier nucleation and accelerated percolation of reversed domains across the
	film thickness. Symmetric three-layer architectures with an interior AFE-rich
	layer maximize this cooperative effect, whereas asymmetric stacks yield
	intermediate trade-offs.
	
	These results help in establishing clear design principles for FE--AFE hafnia
	heterostructures under fixed thickness and identical electrical loading: vertical
	phase segregation enables substantial coercive-field reduction without severe
	remanence loss, interior placement of moderately AFE-rich layers is optimal, and
	multilayer architectures can be exploited to engineer switching behavior beyond
	what is achievable in single-layer polycrystalline films. Thus, the proposed model provides a predictive platform for phase-engineered hafnia-based ferroelectric capacitors.
	
        Beyond its immediate application to ferroic switching physics,
		the present phase-field framework provides a physics-consistent basis for
		generating high-fidelity datasets describing polarization evolution,
		domain interactions, and microstructure-dependent switching behavior in
		mixed-phase hafnia thin films. Such datasets can enable the development of
		data-driven surrogate models, including graph neural networks~\cite{KEVIN} and
		neural-operator approaches~\cite{li2020fourier}, for accelerated exploration of
		microstructure–property relationships. Detailed results from these efforts
		will be reported in future studies.
	
	\section*{Declaration of Competing Interest}
	The authors declare that they have no known competing financial interests or personal relationships that could have appeared to influence the work reported in this paper.
	
\section*{Acknowledgments}
This work was supported by the National Research Foundation of Korea (NRF) funded by the Ministry of Science and ICT through the Nano \& Material Technology Development Program (RS-2024-00444182) and another NRF grant (RS-2024-00450836).

	\bibliographystyle{elsarticle-num}
	\bibliography{References}
	
\end{document}

\begin{center}
	{\LARGE \textbf{Supplementary Information For}}\\[6pt]
	
	{\large ``Unified Phase-Field Framework for Antiferroelectric, Ferroelectric and Dielectric Phases: Application to HZO Thin Films''}\\[8pt]
	
	{\large
		P.~Pankaj$^{1}$,
		Sandeep Sugathan$^{1}$,
		Si~Joon~Kim$^{2}$,
		Pil-Ryung Cha$^{1,*}$
	}\\[4pt]
	
	{\small
		$^{1}$School of Materials Science and Engineering, Kookmin University, Seoul 02707, Republic of Korea\\
		$^{2}$Department of Electrical and Electronics Engineering, Kangwon National University, Chuncheon, Gangwon-Do 24341, Republic of Korea\\[4pt]
		$^{*}$Corresponding author: cprdream@kookmin.ac.kr
	}
\end{center}
	
	\vspace{1em}
	
	\section{Nondimensionalization of the Free-Energy Functional and TDGL equations}
The phase-field simulations are formulated in terms of the total free-energy
functional (Eq.~1) and the corresponding TDGL equations (Eq.~10). To facilitate
numerical implementation, the governing equations are rewritten in nondimensional
form by introducing characteristic reference scales for polarization, free-energy
density, length, time, and electric field. The polarization fields are scaled as
$P_m = P_0 P_m'$ and $P_s = P_0 P_s'$, the free-energy density as $f = f_0 f'$,
spatial coordinates as $\mathbf{r} = l_0 \mathbf{r}'$, and time as $t = t_0 t'$.

In the present work, the characteristic length and time scales are chosen as
$l_0 = 1~\mathrm{nm}$ and $t_0 = 3~\mathrm{ns}$, consistent with the domain-wall
thickness and intrinsic polarization relaxation times in hafnia-based ferroelectric
and antiferroelectric thin films. The free-energy density scale is defined as
$f_0 = \tfrac{1}{4}\alpha_0 P_0^2$, where $\alpha_0$ is a reference Landau stiffness,
and the electric field is scaled as $\mathbf{E} = E_0 \mathbf{E}'$ with
$E_0 = f_0 / P_0$.

The characteristic polarization scale $P_0$ is chosen as the spontaneous
polarization magnitude corresponding to the minimum of the bulk
free-energy density in the Kittel-type two-sublattice formulation.
For the sixth-order polynomial expansion used here, this yields
\begin{equation}
	P_0^2
	=
	\frac{2|\beta| + 2\sqrt{\beta^2 - 4\gamma(\alpha-g)}}{\gamma}.
\end{equation}
For the parameters used in this work, $P_0 = 41 \mu$C/cm$^2$.

\subsection{Bulk free energy density}

We consider the nondimensional form of the local bulk free-energy density.
Substituting the scaling relations introduced above into Eq.~(4), the nondimensional
bulk free-energy density in the local frame is obtained as

\begin{equation}
	\begin{aligned}
		f_{\mathrm{bulk}}'^L &=
		\alpha'\big[(P_{m,z}'^L)^2 + (P_{s,z}'^L)^2\big]
		+ \beta'\big[(P_{m,z}'^L)^4 + 6 (P_{m,z}'^L)^2 (P_{s,z}'^L)^2 + (P_{s,z}'^L)^4\big] \\
		&\quad + \gamma'\big[(P_{m,z}'^L)^6 + 15 (P_{m,z}'^L)^4 (P_{s,z}'^L)^2
		+ 15 (P_{m,z}'^L)^2 (P_{s,z}'^L)^4 + (P_{s,z}'^L)^6\big] \\
		&\quad + g'\big[(P_{m,z}'^L)^2 - (P_{s,z}'^L)^2\big]
		+ a_0'\big[(P_{m,x}'^L)^2 + (P_{m,y}'^L)^2
		+ (P_{s,x}'^L)^2 + (P_{s,y}'^L)^2\big],
	\end{aligned}
	\label{eq:A6_bulk_nd}
\end{equation}
where the nondimensional Landau coefficients are defined as
\begin{equation*}
	\alpha' = \frac{\alpha(T)}{\alpha_0}, \quad
	\beta' = \frac{\beta P_0^2}{8\,\alpha_0}, \quad
	\gamma' = \frac{\gamma P_0^4}{48\,\alpha_0}, \quad
	g' = \frac{g}{\alpha_0}, \quad
	a_0' = \frac{2}{\chi_f\,\alpha_0}.
\end{equation*}

\subsection{Gradient energy density}

The nondimensional gradient energy density is written as:
\begin{equation}
	f_{\mathrm{grad}}'^G =
	G_{11}'
	\sum_{i=x,y,z}
	\left[
	(\nabla' P_{m,i}'^G)^2
	+
	(\nabla' P_{s,i}'^G)^2
	\right].
\end{equation}
Here, the nondimensional gradient coefficient is defined as
$G_{11}' = 2 G_{11} / (\alpha_0 l_0^2)$.

\subsection{Applied electric field energy density}

The nondimensional applied field energy density is written as
\begin{equation}
	f_{\mathrm{ap}}'^{\,G} = - E_{\mathrm{ap},i}'^{\,G}\, P_{m,i}'^{\,G},
	\label{eq:app_field_nd}
\end{equation}
where the nondimensional applied field is defined as
\begin{equation*}
	E_{\mathrm{ap},i}'^{\,G} = \frac{E_{\mathrm{ap},i}^{G}}{E_0}.
\end{equation*}
For out-of-plane loading along the $z$ axis, Eq.~\eqref{eq:app_field_nd} reduces to
$f_{\mathrm{ap}}'^{\,G} = -E_{\mathrm{ap},z}'^{\,G} P_{m,z}'^{\,G}$.

\subsection{Electrostatic energy density and Poisson equation}

The nondimensional electrostatic energy density is written as
\begin{equation}
	f_{\mathrm{elc}}'^{\,G}
	=
	-\tfrac{1}{2}\, E_i'^{\,G} P_{m,i}'^{\,G},
	\label{eq:felc_nd}
\end{equation}
where $E_i'^{\,G}=E_i^{G}/E_0$.

The electrostatic potential is obtained by solving the nondimensional Poisson
equation (Eq.~(7))
\begin{equation}
	(\varepsilon_{ij}\phi'_{,j})_{,i}
	=
	\mathrm{C}\,P'_{m,i,i},
	\qquad
	\mathrm{C}=\frac{l_0 P_0}{\varepsilon_0\phi_0},
	\label{eq:poisson_nondim_code}
\end{equation}
where, $\phi_0 = 4E_0 l_0$. The dimensionless internal
electric field is then computed as $E'_i=-\phi'_{,i}$.

\subsection{Nondimensional evolution equations}
Using the nondimensional free-energy densities defined in the previous section,
the TDGL equations governing the evolution of the macroscopic and staggered
polarization fields can be written in nondimensional form as
\begin{equation}
	\frac{\partial P_{m,i}'^{\,G}}{\partial t'}
	= 	- L'
	\left[
	\frac{\partial f_{\mathrm{bulk}}'}{\partial P_{m,i}'^{\,G}}
	+ \frac{\partial f_{\mathrm{elc}}'}{\partial P_{m,i}'^{\,G}}
	+ \frac{\partial f_{\mathrm{ap}}'}{\partial P_{m,i}'^{\,G}}
	- \nabla' \cdot
	\frac{\partial f_{\mathrm{grad}}'}{\partial \nabla' P_{m,i}'^{\,G}}
	\right],
	\label{eq:TDGL_Pm_nd}
\end{equation}
and
\begin{equation}
	\frac{\partial P_{s,i}'^{\,G}}{\partial t'}
	=
	- L'
	\left[
	\frac{\partial f_{\mathrm{bulk}}'}{\partial P_{s,i}'^{\,G}}
	- \nabla' \cdot
	\frac{\partial f_{\mathrm{grad}}'}{\partial \nabla' P_{s,i}'^{\,G}}
	\right],
	\label{eq:TDGL_Ps_nd}
\end{equation}
where the nondimensional kinetic coefficient is defined as
\begin{equation}
	L' = \frac{L f_0 t_0}{P_0^2}.
\end{equation}

\section{Electric-field sweep rate and time discretization}

To reproduce experimentally relevant switching conditions, a triangular electric-field
waveform with frequency $\nu = 10~\mathrm{kHz}$ and peak amplitude
$E_m = 3~\mathrm{MV/cm}$ is applied across a film of thickness $t_z = 10~\mathrm{nm}$,
corresponding to a voltage amplitude $V_m = E_m t_z = 3~\mathrm{V}$. For a full triangular
cycle, the electric-field ramp rate is
\[
\frac{dE}{dt} = 4E_m \nu = 1.2\times10^{5}~\mathrm{MV/(cm\cdot s)}.
\]

The applied field is incremented in discrete steps of
$\Delta E = 0.15~\mathrm{MV/cm}$, yielding a physical time interval per step of
\[
\Delta t_{\mathrm{phys}} = \frac{\Delta E}{dE/dt} = 1.25~\mu\mathrm{s}.
\]

Within each field increment, the TDGL equations are integrated using a fixed physical
timestep of $t_0 = 3~\mathrm{ns}$, resulting in approximately
\[
N_s = \frac{\Delta t_{\mathrm{phys}}}{t_0} \approx 416
\]
TDGL substeps per field update. This discretization preserves the experimentally relevant
sweep rate while ensuring sufficient temporal resolution to capture intrinsic
polarization relaxation dynamics.

All simulations are performed in nondimensional form with spatial discretization
$\Delta h = 1$ and temporal discretization $\Delta t = 0.02$. Unless otherwise noted, all polarization microstructure figures use identical color scales expressed in $\mu$C/cm$^2$.

\section{Polarization microstructures for single crystal}

\begin{figure}
	\centering
	\includegraphics[width=0.8\textwidth]{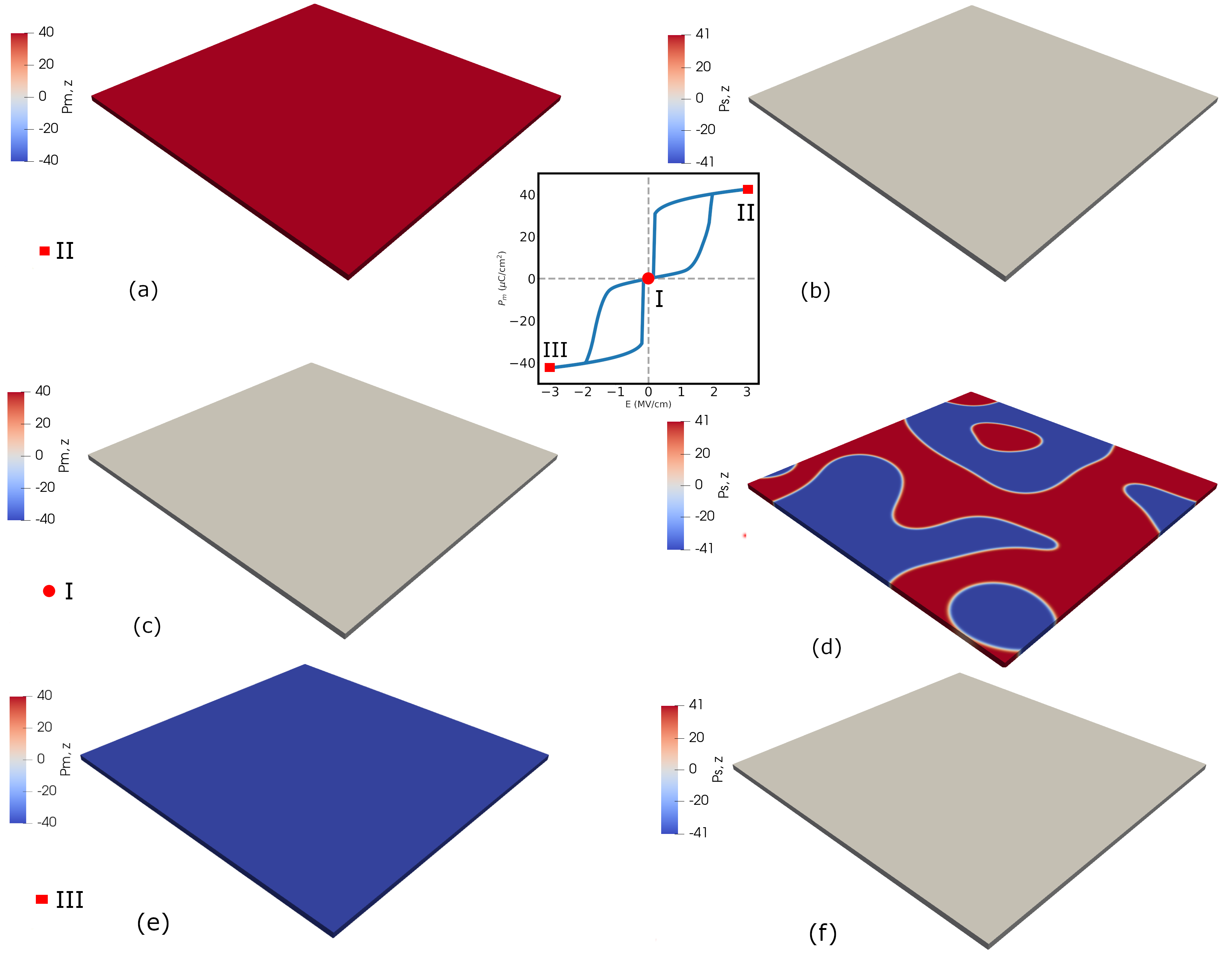}
\caption{Spatial distributions of $P_{m,z}$ and $P_{s,z}$ in the single-crystal
	AFE film at three characteristic points on the $P$--$E$ curve
	corresponding to the intermediate sweep rate (green curve in Fig.~3):
	(I) $E=E_{\max}$, (II) $E=0$, and (III) $E=-E_{\max}$.
	The panels show the field-induced FE state, re-emergence of antipolar
	domains at zero field, and the reversed FE state, respectively. 
}
	\label{fig:_SI_singleX}
\end{figure}

Fig.~\ref{fig:_SI_singleX} shows the spatial distributions of $P_{m,z}$ and $P_{s,z}$ in the single-crystal AFE film at three characteristic applied electric field states corresponding to the
intermediate sweep-rate (green) trajectory in Fig.~3: $E=E_{\max}$, $E=0$, and
$E=-E_{\max}$.

At $E=E_{\max}$, the system adopts a uniform field-induced FE state
characterized by finite $P_{m,z}$ and vanishing $P_{s,z}$, indicating suppression
of antipolar order under strong applied electric field. Upon removal of the field
($E=0$), $P_{m,z}$ collapses toward zero while antipolar order re-emerges through
the formation of domains with opposite staggered polarization
($+P_{s,z}$ and $-P_{s,z}$) separated by diffuse domain walls. 

At $E=-E_{\max}$, the polarization reverses to the oppositely oriented
FE state with nearly uniform negative $P_{m,z}$ and suppressed
$P_{s,z}$. These field-resolved microstructures illustrate how kinetic effects
govern the transient evolution between antipolar and polar states even in the
absence of microstructural disorder, providing a microstructural basis for the
deviations from the quasi-static thermodynamic response discussed in the main
text.

	\section{Polarization microstructures for single-layer 100\% FE film}

\begin{figure}[t]
	\centering
	\includegraphics[width=0.7\textwidth]{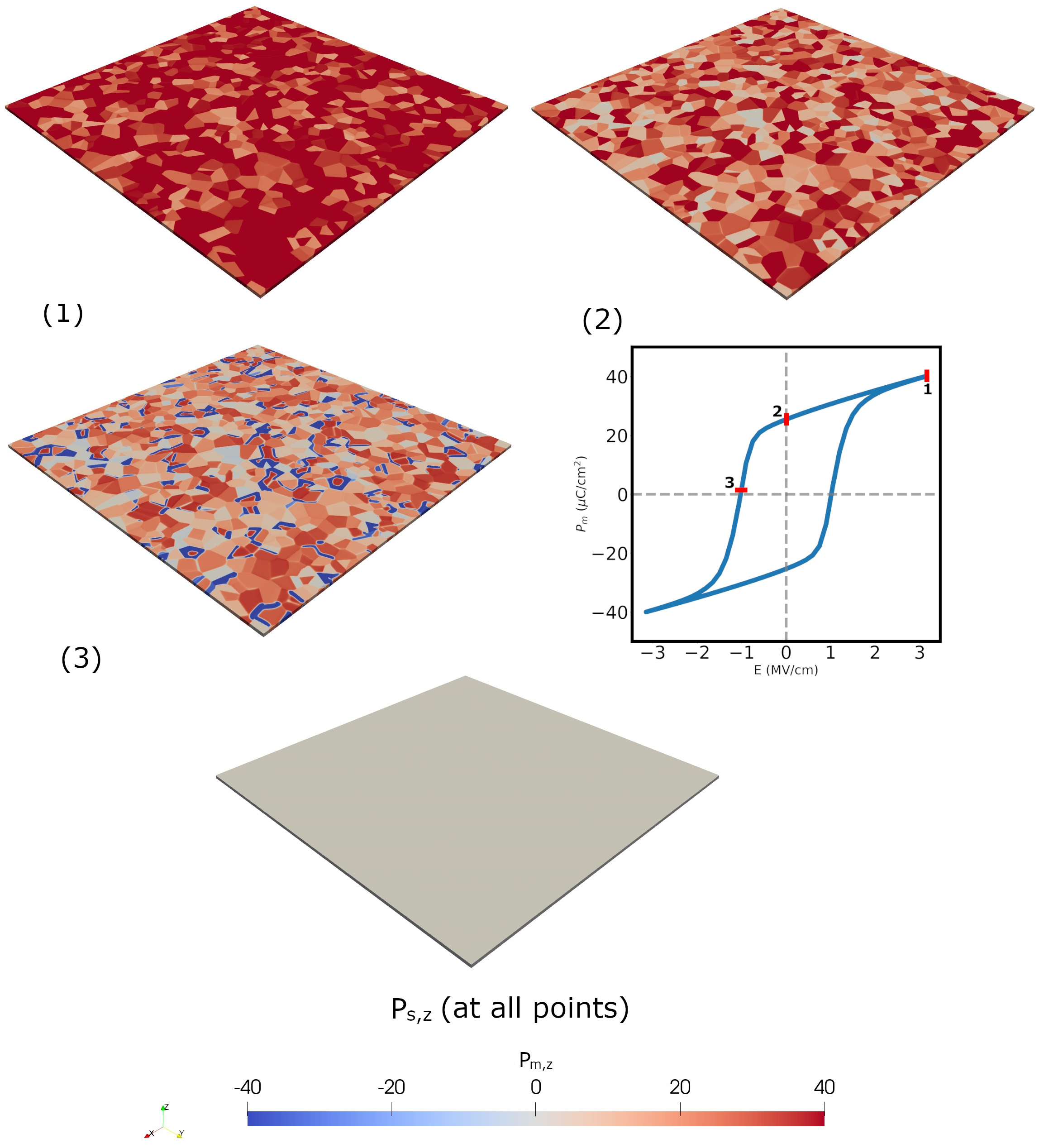}
	\caption{Grain-resolved microstructures of $P_{m,z}$ in the single-layer 100\%
		FE film at three characteristic points on the $P$--$E$ curve:
		(1) $E=+E_{\max}$, (2) $E=0$ after field reversal, and
		(3) the pre-switching state on the reverse branch just before
		polarization reversal ($E\lesssim E_c$).
		}
	\label{fig:SI_microstructures_FE}
\end{figure}

Fig.~\ref{fig:SI_microstructures_FE} presents grain-resolved polarization microstructures of the single-layer 100\% FE film, shown in Fig.~2(a), at three characteristic points marked along the $P$--$E$ loop. At the maximum applied electric field ($E=+E_{\max}$), the microstructure is dominated by a uniformly positive $P_{m,z}$ across nearly all grains (Fig.~\ref{fig:SI_microstructures_FE}(1)), corresponding to a field-stabilized FE state. While the magnitude of $P_{m,z}$ varies from grain to grain due to crystallographic misorientation, no antipolar domain features are observed.

Upon field reversal to zero field ($E=0$), $P_{m,z}$ decreases from its saturated value but remains positive in most grains (Fig.~\ref{fig:SI_microstructures_FE}(2)), indicating a finite $P_r$ inherited from the preceding field-stabilized state. Compared to the saturated configuration, the microstructure exhibits increased spatial heterogeneity, reflecting grain-scale variations in switching response.

Just prior to polarization reversal ($E \lesssim E_c$), $P_{m,z}$ becomes strongly heterogeneous (Fig.~\ref{fig:SI_microstructures_FE}(3)), with coexistence of positively polarized grains and localized regions of reversed polarization. This microstructure reflects the onset of distributed grain-scale switching preceding global reversal. 

Throughout the entire field cycle, $P_{s,z}$ remains identically zero, confirming that switching in the 100\% FE film is governed exclusively by $P_{m,z}$, in contrast to the antipolar-domain-mediated behavior observed in AFE films.

	\section{Grain-Resolved Origin of Residual Antipolar Order}
\begin{figure}[ht]
		\centering
		\includegraphics[width=\linewidth]{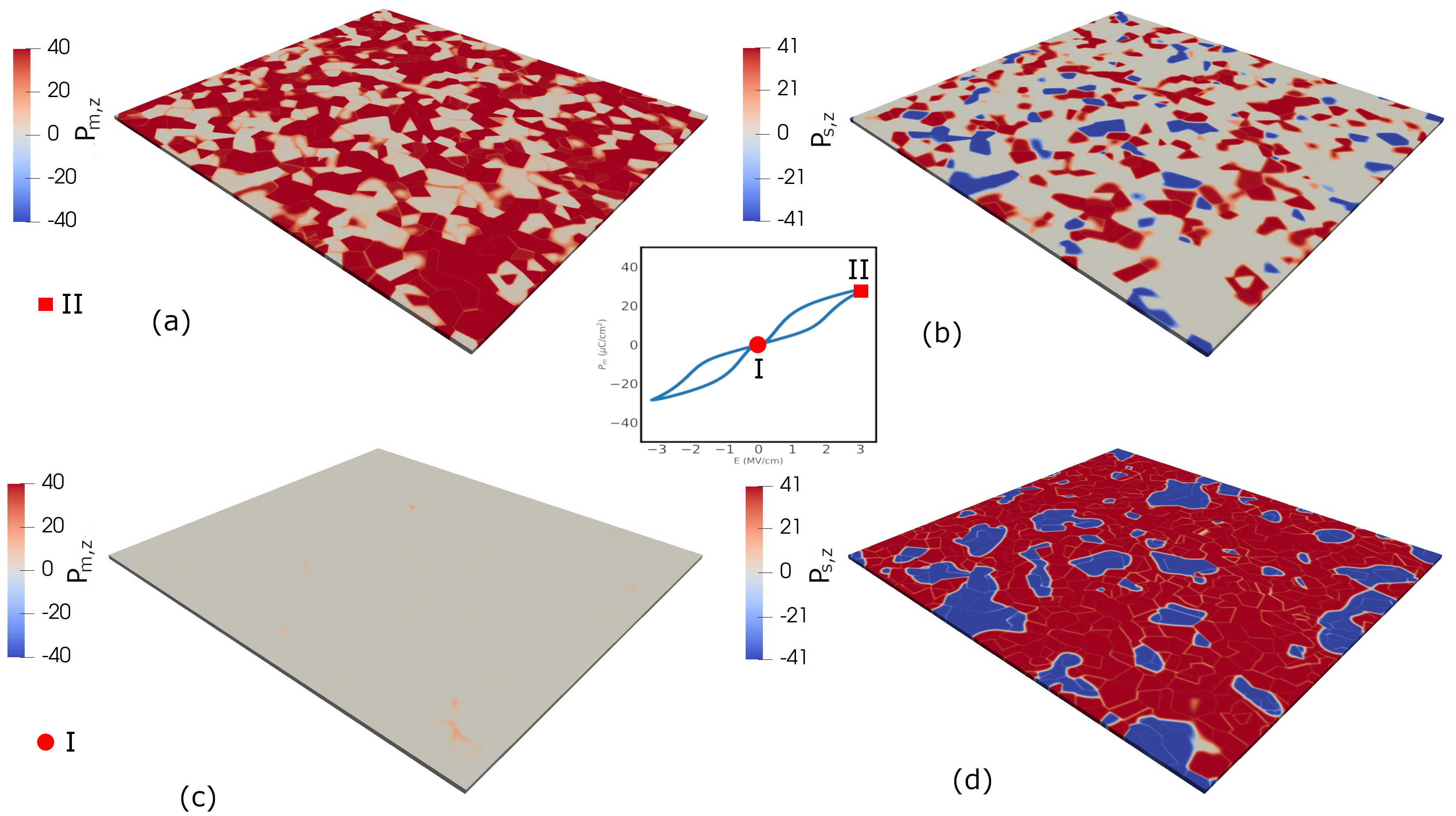}
		\caption{
			Local-frame polarization microstructures corresponding to the global
			fields shown in Fig.~5 for a 100\% AFE single-layer polycrystalline film.
			Panels (a,b) show $P_{m,z}^L$ and $P_{s,z}^L$ at $E = +E_{\max}$, while panels
			(c,d) show the corresponding fields at $E = 0$.
			At $E=+E_{\max}$, $P_{m,z}^L$ approaches saturation in most grains, while
			$P_{s,z}^L$ exhibits strong grain-to-grain variability, collapsing in some grains but remaining finite in others.
			This heterogeneous local response explains why $P_{s,z}^G$ does not vanish
			completely under strong applied fields.
			At $E = 0$, $P_{s,z}^L$ recovers large-magnitude antipolar domains.
			Comparison of panel (d) with Fig. 5(d) shows that
			$P_{s,z}^G$ reflects spatial averaging of locally saturated antipolar states.
		}
		\label{fig:SI_local_microstructures_AFE}
	\end{figure}

	Fig.~\ref{fig:SI_local_microstructures_AFE} provides a grain-resolved, local-frame view of the polarization fields
	corresponding to the global microstructures shown in Fig.~5. At
	$E = +E_{\max}$, panel (b) reveals that while $P_{s,z}^L$ collapses in many grains, it remains finite in a subset of grains due
	to crystallographic misorientation and incomplete coupling to the applied field.
	As a result, spatial averaging over heterogeneous local responses yields a
	nonzero but reduced global $P_{s,z}^G$, even though a fully suppressed antipolar
	state might be expected under strong electric fields.
	
	At $E = 0$, both local- and global-frame representations show recovery of
	well-defined antipolar domains, with $P_{m,z} \approx 0$ throughout the film.
	The local-frame fields confirm that the restored global $P_{s,z}^G$ arises from
	locally saturated antipolar states rather than uniform weak ordering. These
	observations clarify the microscopic origin of the residual antipolar signal
	under field and the reversible recovery of antipolar order upon unloading.
	
	% ---- Figure S1 (your local-frame figure) ----
	
	\section{Extracted Switching Metrics}
	
Tables~\ref{tab:PrEc_baseline1} and \ref{tab:PrEc_baseline2} summarize the
numerical values of the average remanent polarization $P_{r,\mathrm{avg}}$
and coercive field $E_{c,\mathrm{avg}}$ used in Fig.~9. For each
architecture, switching metrics are extracted from four independent
grain-topology realizations, and the reported values correspond to the
ensemble mean together with one standard deviation (STD), providing a
quantitative measure of variability associated with stochastic
microstructure generation.
	
\begin{table}[htbp]
	\centering
	\caption{Extracted switching metrics for Baseline~1
		(AFE~10\%–FE~90\%) polycrystalline films.
Values represent mean $\pm$ standard deviation over four
independent grain-topology realizations ($n=4$).}
	\vspace{5pt}
	\label{tab:PrEc_baseline1}
	
	\begin{tabular}{lcccc}
		\hline
		Architecture & $P_{r,\mathrm{avg}}$ & STD &
		$E_{c,\mathrm{avg}}$ & STD \\
		& ($\mu$C/cm$^2$) & ($\mu$C/cm$^2$)
		& (MV/cm) & (MV/cm)  \\
		\hline
		1L    & 23.424 & 0.335 & 0.983 & 0.011 \\
		1-9   & 22.976 & 0.653 & 0.878 & 0.007 \\
		2-8   & 22.058 & 0.916 & 0.836 & 0.010 \\
		5-5   & 19.936 & 0.391 & 0.712 & 0.008 \\
		4-2-4 & 23.348 & 0.332 & 0.718 & 0.015 \\
		2-6-2 & 19.801 & 0.698 & 0.632 & 0.011 \\
		\hline
	\end{tabular}
\end{table}
	
	\begin{table}[htbp]
		\centering
		\caption{Extracted switching metrics for Baseline~2
			(AFE~30\%–FE~70\%) polycrystalline films.
Values represent mean $\pm$ standard deviation (STD) over four
independent grain-topology realizations ($n=4$).}
		\vspace{5pt}
		\label{tab:PrEc_baseline2}
		
		\begin{tabular}{lcccc}
			\hline
			Architecture & $P_{r,\mathrm{avg}}$ & STD &
			$E_{c,\mathrm{avg}}$ & STD  \\
			& ($\mu$C/cm$^2$) & ($\mu$C/cm$^2$)
			& (MV/cm) & (MV/cm)  \\
			\hline
			1L    & 19.441 & 0.419 & 0.917 & 0.021  \\
			1-9   & 19.359 & 0.614 & 0.823 & 0.003  \\
			2-8   & 18.495 & 0.685 & 0.785 & 0.007  \\
			5-5   & 17.378 & 0.197 & 0.679 & 0.006  \\
			4-2-4 & 20.620 & 0.288 & 0.658 & 0.014  \\
			2-6-2 & 17.435 & 0.407 & 0.621 & 0.010  \\
			\hline
		\end{tabular}
	\end{table}